\documentclass[aos,MSNbibl,dvips]{arximspdf}
\usepackage{graphicx}
\doi{10.1214/13-AOS1138} 
\volume{41}
\issue{4}
\pubyear{2013}
\firstpage{2097}
\lastpage{2122}

\makeatletter

\newtheorem{lem}{Lemma}
\newtheorem{theorem}{Theorem}

\newproclaim{rem}{Remark}

\newcommand{\implies}{\Longrightarrow}

\newcommand{\ex}{\mathbb{E}}

\newcommand{\bdot}{\bolds\cdot}

\newcommand{\bb}{\mathbf{b}}
\newcommand{\PL}{\mathrm{PL}}
\newcommand{\CPL}{\mathrm{CPL}}
\newcommand{\pr}{\mathbb{P}}
\newcommand{\pih}{\hat{\pi}}
\newcommand{\Ph}{\hat{P}}
\newcommand{\Lamh}{\hat{\Lambda}}
\newcommand{\thetah}{\hat\theta}
\newcommand{\prpl}{\pr_{\mathrm{PL}}}
\newcommand{\prcpl}{\pr_{\mathrm{CPL}}}
\newcommand{\diag}{\operatorname{diag}}
\newcommand{\reals}{\mathbb{R}}

\newcommand{\Cc}{\mathcal{C}}
\newcommand{\Sc}{\mathcal{S}}
\newcommand{\Ec}{\mathcal{E}}
\newcommand{\ch}{\hat{c}}
\newcommand{\eps}{\varepsilon}
\newcommand{\var}{\operatorname{var}}
\newcommand{\Pt}{\widetilde{P}}
\newcommand{\ber}{\operatorname{Ber}}
\newcommand{\ah}{\hat{a}}
\newcommand{\bh}{\hat{b}}
\newcommand{\ppb}{\bar{p}}
\newcommand{\Pc}{\mathcal{P}}
\newcommand{\agam}{a_\gamma}
\newcommand{\Ais}{\Ad_{i*}}
\newcommand{\Asi}{\Ad_{*i}}
\newcommand{\Qu}{\widetilde{Q}}
\newcommand{\qqb}{\bar{q}}
\newcommand{\bd}{\widetilde{b}}
\newcommand{\xid}{\widetilde{\xi}}
\newcommand{\Nisd}{\widetilde{N}}
\newcommand{\Ad}{\widetilde{A}}
\newcommand{\Pd}{\widetilde{P}}
\newcommand{\Misd}{\widetilde{M}}
\newcommand{\mub}{\bar{\mu}}

\newcommand{\gam}{\gamma}
\newcommand{\permute}{\phi}

\makeatother

\begin{document}
\begin{frontmatter}

\title{Pseudo-likelihood methods for community detection in large
sparse networks\thanksref{T1}}
\runtitle{Pseudo-likelihood methods for community detection}

\thankstext{T1}{Supported by the NSF Focused Research Group Grant DMS-11-59005.}

\begin{aug}
\author[A]{\fnms{Arash A.} \snm{Amini}\ead [label=earash]{aaamini@umich.edu}},
\author[B]{\fnms{Aiyou} \snm{Chen}\ead[label=eaiyou]{aiyouchen@google.com}},
\author[C]{\fnms{Peter J.} \snm{Bickel}\ead[label=epeter]{bickel@stat.berkeley.edu}}\\
\and
\author[A]{\fnms{Elizaveta} \snm{Levina}\corref{}\thanksref{t2}\ead[label=eliza]{elevina@umich.edu}}

\runauthor{Amini, Chen, Bickel and Levina}
\affiliation{University of Michigan,
Google, Inc., University of California, Berkeley, and~University of Michigan}
\address[A]{A. A. Amini\\
E. Levina\\
Department of Statistics\\
University of Michigan\\
Ann Arbor, Michigan 48109-1107\\
USA\\
\printead{earash}\\
\hphantom{E-mail: }\printead*{eliza}}
\address[B]{A. Chen\\
Google, Inc. \\
1600 Amphitheatre Pkwy \\
Mountain View, California 94043 \\
USA\\
\printead{eaiyou}}
\address[C]{P. J. Bickel\\
Department of Statistics \\
University of California, Berkeley \\
Berkeley, California 94760 \\
USA\\
\printead{epeter}} 
\end{aug}

\thankstext{t2}{Supported by NSF Grant DMS-11-06772.}

\received{\smonth{12} \syear{2012}}
\revised{\smonth{5} \syear{2013}}

%
\begin{abstract}
Many algorithms have been proposed for fitting network models with
communities, but most of them do not scale well to large networks, and
often fail on sparse networks. Here we propose a new fast
pseudo-likelihood method for fitting the stochastic block model for
networks, as well as a variant that allows for an arbitrary degree
distribution by conditioning on degrees. We show that the algorithms
perform well under a range of settings, including on very sparse
networks, and illustrate on the example of a network of political
blogs. We also propose spectral clustering with perturbations, a
method of independent interest, which works well on sparse networks
where regular spectral clustering fails, and use it to provide an
initial value for pseudo-likelihood. We prove that pseudo-likelihood
provides consistent estimates of the communities under a mild condition
on the starting value, for the case of a block model with two
communities.
\end{abstract}

%

\end{frontmatter}

\section{Introduction}\label{secintro}
Analysis of network data is important in a range of disciplines and
applications, appearing in such diverse areas as sociology,
epidemiology, computer science, and national security, to name a few.
Network data here refers to observed edges between nodes, possibly
accompanied by additional information on the nodes and/or the edges,
for example, edge weights. One of the fundamental questions in analysis
of such
data is detecting and modeling community structure within
the network. A lot of algorithmic approaches to community detection
have been proposed, particularly in the physics literature; see
\cite{NewmanPNAS,Fortunato2010} for reviews. These include
various greedy methods such as hierarchical clustering (see \cite
{Newman2004Review} for a review)
and algorithms based on optimizing a global criterion over all
possible partitions, such as normalized
cuts \cite{Shi00} and modularity \cite{Newman&Girvan2004}. The
statistics literature has been more
focused on model-based methods, which postulate and fit a
probabilistic model for a network with communities. These include
the popular stochastic block model \cite{Holland83}, its
extensions to include varying degree distributions within communities
\cite{Karrer10}
and overlapping
communities \cite{Airoldi2008,Hall&Karrer&Newman2011}, and various
latent variable models \cite{Handcock2007,Hoff2007}.

The stochastic block model is perhaps the most commonly used and best
studied model for community detection. For a network with $n$ nodes
defined by its $n \times n$ adjacency matrix $A$, this model
postulates that the true node labels
$c=(c_{1},\ldots,c_{n})\in\{1,\ldots,K\}^{n}$ are drawn
independently from the multinomial distribution with parameter $\pi
=(\pi_{1},\ldots,\pi_{K})$, where $\pi_i >0$ for all $i$, and
$K$ is the number of communities,
assumed known. Conditional on the labels, the edge variables $A_{ij}$
for $i < j$ are independent Bernoulli variables with
%
\begin{equation}
\label{bm} \ex[A_{ij}|c] = P_{c_i c_j},
\end{equation}
where $P=[P_{ab}]$ is a $K \times K$ symmetric matrix. The network
is undirected, so $A_{ji} = A_{ij}$, and $A_{ii}=0$ (no self-loops).
The problem of community detection is then to infer the node labels
$c$ from $A$, which typically also involves estimating $\pi$ and $P$.

There are many extensions of the block model, notably to mixed
membership models \cite{Airoldi2008}, but we will only focus on
one extension here that we use later in the paper. The block model implies
the same expected degree for all nodes within a community,
which excludes networks with ``hub'' nodes commonly encountered in
practice. The degree-corrected block model \cite{Karrer10} removes
this constraint by replacing (\ref{bm}) with $\ex[A_{ij}|c] = \theta
_i \theta_j P_{c_i c_j}$,
where $\theta_i$'s are node degree parameters which satisfy an
identifiability constraint. If the degree parameters only take on a
discrete number of values, one can think of the degree-corrected
block model as a regular block model with a larger number of blocks,
but that
loses the original interpretation of communities.
In \cite{Karrer10} the Bernoulli
distribution for $A_{ij}$ was replaced by the Poisson, primarily for
ease of technical derivations, and in fact this is a good
approximation for a range of networks \cite{Perry2012}.

Fitting block models is nontrivial, especially for large networks,
since in principle the problem of optimizing over all possible label
assignments is NP-hard. In the Bayesian framework, Markov Chain Monte
Carlo methods have been
developed \cite{Snijders&Nowicki1997,Nowicki2001}, but they only
work for networks with a few hundred nodes. Variational methods
have also been developed and studied (see, e.g.,
\cite
{Airoldi2008,Celisseetal2011,Mariadassouetal2010,Bickel&Choi&etal2012}),
and are generally
substantially faster than
the Gibbs sampling involved in MCMC, but still do not scale to the
order of a million nodes. Another Bayesian approach based on a belief
propagation algorithm
was proposed recently by Decelle et al. \cite{Decelleetal2011}, and
is comparable to ours in
theoretical complexity, but slower in practice; see more on this in
Section~\ref{secsimulations}.\looseness=1

In the non-Bayesian framework, a profile likelihood approach was proposed
in \cite{Bickel&Chen2009}: since for a given
label assignment parameters can be estimated trivially by plug-in, they
can be profiled out and the resulting criterion
can be maximized over all label assignments by greedy search. The
same method is used in \cite{Karrer10} to fit the degree-corrected
block model. The speed of the profile likelihood algorithms depends on
exactly what
search method is used and the number of iterations it is run for, but
again these
generally work well for thousands but not millions of nodes. A
method of moments approach was proposed in
\cite{Bickel&Chen&Levina2011}, for a large class of network models
that includes the block model as a special case. The generality of
this method is an advantage, but it involves counting all occurrences
of specific patterns in the graph, which is computationally
challenging beyond simple special cases. Some faster approximations
for block model fitting based on spectral representations are also available
\cite{Newman2006,Rohe2011}, but the properties of
these approximations are only partially known.

Profile likelihood methods have been proven to
give consistent estimates of the labels when the degree of the graph
grows with the number of nodes, under both the stochastic block models
\cite{Bickel&Chen2009} and the degree-corrected version~\cite{Zhaoetal2012}.
To obtain ``strong consistency'' of the labels,
that is, the probability of the estimated label vector being equal to
the truth converging to 1, the average graph degree $\lambda_n$ has to
grow faster
than $\log n$, where $n$ is the number of nodes. To obtain ``weak
consistency,'' that is, the fraction of misclassified nodes converging to
0, one only needs $\lambda_n \rightarrow\infty$.
Asymptotic behavior of variational methods is studied in
\cite{Celisseetal2011} and \cite{Bickel&Choi&etal2012}, and in
\cite{Decelleetal2011} this belief propagation method is analyzed for
both the sparse [$\lambda_n = O(1)$] and the dense ($\lambda_n
\rightarrow\infty$) regimes, by nonrigorous cavity methods from
physics, and a phase transition threshold, below which the
labels cannot be recovered, is established. In fact, it is easy to see that
consistency is impossible to achieve unless $\lambda_n \rightarrow
\infty$, since otherwise the expected fraction of isolated nodes does
not go to 0. The results one can get for the sparse case, such as~\cite
{Decelleetal2011}, can only
claim that the estimated labels are correlated with the truth better
than random guessing, but not that they are consistent. In this
paper, for the purposes of theory we focus on consistency and thus
necessarily assume that the degree grows with~$n$. However, in
practice we find that our methods are very well suited for sparse
networks and work well on graphs with quite small degrees.

Our main contribution here is a new fast pseudo-likelihood algorithm for
fitting the block model, as well as its variation conditional on node
degrees that allows for fitting networks with highly variable node
degrees within communities. The idea of pseudo-likelihood dates back
to \cite{Besag74}, and in general amounts to ignoring some of the
dependency structure of the data in order to simplify the
likelihood\vadjust{\goodbreak}
and make it more tractable. The main feature of the adjacency matrix
we ignore here is its symmetry; we also apply block compression, that
is, divide the nodes into blocks and only look at the likelihood of
the row sums within blocks. This leads to an accurate and
fast approximation to the block model likelihood, which allows us to
easily fit block models to networks with tens of millions of nodes.
Another major contribution of the paper is the consistency proof of
one step of the algorithm. The proof requires new and somewhat
delicate
arguments not previously used in consistency proofs for networks; in
particular, we use the device of assuming an initial value that has a
certain overlap with the truth, and then show the amount of overlap
can be arbitrarily close to purely random. Finally, we propose
spectral clustering with perturbations, a new clustering method of independent
interest which we use to
initialize pseudo-likelihood in practice. For sparse networks, regular
spectral clustering often
performs very poorly, likely due to the presence of many disconnected
components. We perturb the network by adding additional weak edges to
connect these components, resulting in regularized spectral clustering
which performs well under a wide range of settings.

The rest of the paper is organized as follows. We present the
algorithms in Section \ref{secalgorithms}, and prove asymptotic
consistency of pseudo-likelihood in Section~\ref{secconsist}. The
numerical performance of the methods is
demonstrated on a
range of simulated networks in Section \ref{secsimulations} and on a
network of political
blogs in Section \ref{secexample}. Section \ref{secdiscuss}
concludes with discussion, and the \hyperref[app]{Appendix} contains some additional
technical results.

\section{Algorithms}
\label{secalgorithms}

\subsection{Pseudo-likelihood}
The joint likelihood of $A$ and $c$ could in principle be maximized
via the expectation--maximization (EM) algorithm,
but the E-step involves optimizing over all possible label assignments,
which is NP-hard. Instead, we introduce an initial labeling
vector $e=(e_{1},\ldots,e_{n})$, $e_{i} \in\{1,\ldots,K\}$, which
partitions the nodes into $K$
groups. Note that
for convenience we partition into the same number of groups as we
assume to exist in the true model, but in principle the same idea
can be applied with a different number of groups; in fact dividing
the nodes into $n$ groups with a single node in each group instead
gives an algorithm equivalent to that of~\cite{Newman&Leicht2007}.

The main quantity we work with are the block sums along the columns,
%
\begin{equation}
\label{eqbikdef} b_{ik}=\sum_{j}A_{ij}1(e_{j}=k)
\end{equation}
for $i=1,\ldots,n$, $k=1,\ldots,K$. Let $\bb_{i}=(b_{i1},\ldots,b_{iK})$.
Further, let $R$ be the $K\times K$ matrix with entries $\{R_{ka}\}$
given by
%
\begin{equation}
\label{eqRdef} R_{ka}=\frac{1}{n}\sum
_{i=1}^{n}1(e_{i}=k,c_{i}=a).
\end{equation}
Let $R_{k\bdot}$ be the $k$th row of $R$, and let $P_{\bdot l}$
be the $l$th column of $P$. Let $\lambda_{lk}=nR_{k\bdot}P_{\bdot l}$
and $\Lambda=\{\lambda_{lk}\}$.

Our approach is based on the following key observations: for each
node $i$, conditional on labels $c=(c_{1},\ldots,c_{n})$ with $c_{i}=l$:
\begin{longlist}[(B)]
\item[(A)] $\{b_{i1},\ldots,b_{iK}\}$ are mutually independent;
\item[(B)] $b_{ik}$, a sum of independent Bernoulli variables,
is approximately Poisson with mean $\lambda_{lk}$.
\end{longlist}
With true labels $\{c_{i}\}$ unknown, each ${\bb}_{i}$ can be viewed
as a mixture of Poisson vectors, identifiable as long as $\Lambda$
has no identical rows.

By ignoring the dependence among $\{\bb_{i},i=1,\ldots,n\}$, using the
Poisson assumption, treating
$\{c_{i}\}$ as latent variables, and setting
$\lambda_{l}=\sum_{k}\lambda_{lk}$,
we can write the pseudo log-likelihood
as follows (up to a constant):
%
\begin{equation}
\label{pl} \ell_{\PL}\bigl(\pi,\Lambda;\{\bb_{i}\}\bigr)=\sum
_{i=1}^{n}\log \Biggl(\sum
_{l=1}^{K}\pi_{l}e^{-\lambda_{l}}\prod
_{k=1}^{K}\lambda _{lk}^{b_{ik}}
\Biggr).
\end{equation}
A pseudo-likelihood estimate of $(\pi,\Lambda)$ can then be obtained
by maximizing $\ell_{\PL}(\pi,\Lambda;\{b_{i}\})$. This can be done
via the standard EM algorithm for mixture models, which
alternates updating parameter values with updating probabilities of
node labels. Once the EM converges, we update the initial block partition
vector $e$ to the most likely label for each node as indicated by
EM, and repeat this process for a fixed number of iterations $T$.

For any labeling $e$, let $n_{k}(e)=\sum_{i}1(e_{i}=k)$,
$n_{kl}(e)=n_{k}(e)n_{l}(e)$ if $k\neq l$, $n_{kk}(e)=n_{k}(e)(n_{k}(e)-1)$
and $O_{kl}(e)=\sum_{i,j}A_{ij}1(e_{i}=k,e_{j}=l)$. We suppress
the dependence on $e$ whenever there is no ambiguity. The details of
the algorithmic steps can be summarized as follows.

\textit{The pseudo-likelihood algorithm.} Initialize labels $e$, and
let $\pih_{l}=n_{l}/n$, $\hat{R}=\operatorname{diag}(\pih_{1},\ldots,\pih_{K})$,
$\Ph_{lk}=O_{lk}/n_{lk}$, $\hat{\lambda}_{lk}=n \hat{R}_{k\bdot
}\Ph_{\bdot l}$, $\Ph= \{\Ph_{lk} \}$ and $\hat{\Lambda} = \{\hat
{\lambda}_{lk} \}$.
Then repeat $T$ times:
\begin{longlist}[{{(6)}}]
\item[{{(1)}}] Compute the block sums $\{b_{il}\}$ according
to (\ref{eqbikdef}).
%

\item[{{(2)}}] Using current parameter estimates $\hat{\pi}$ and
$\hat{\Lambda}$,
estimate probabilities for node labels by
\[
\pih_{il}=\prpl(c_{i}=l | \bb_{i})=
\frac{\pih_{l}\prod_{m=1}^{K}\exp(b_{im}\log\hat{\lambda}_{lm}-\hat{\lambda
}_{lm})}{\sum_{k=1}^{K}\pih_{k}\prod_{m=1}^{K}\exp(b_{im}\log\hat
{\lambda}_{km}-\hat{\lambda}_{km})}.
\]

\item[{{(3)}}] Given label probabilities, update parameter values as
follows:
\[
\pih_{l}=\frac{1}{n}\sum_{i=1}^{n}
\pih_{il},\qquad \hat{\lambda }_{lk}=\frac{\sum_{i}\pih_{il}b_{ik}}{\sum_{i}\pih_{il}}.
\]

\item[{{(4)}}] Return to step 2 unless the parameter estimates have converged.
\item[{{(5)}}] Update labels by $e_{i}=\arg\max_{l}\hat{\pi}_{il}$ and
return to step 1.
\item[{{(6)}}] Update $\Ph$ as follows: 
$\Ph_{lk}= (\sum_{i,j}A_{ij}\pih_{il}\pih_{jk} )/n_{lk}(e)$.
\end{longlist}
In practice, in step 6 we only include the terms corresponding
to $\pih_{il}$ greater than some small threshold. The EM method fits a
valid mixture model as long as the identifiability condition
holds, and is thus guaranteed to converge to a
stationary point of the objective function \cite{Wu1983}.
Another option is to update labels after every parameter update (i.e.,
skip step 4). We have found empirically that the algorithm above
is more stable, and converges faster. In general, we only need a few
label updates until convergence, and even using $T=1$ (one-step label update)
gives reasonable results with a good initial value. The choice of
the initial value of $e$, on the other hand, can be important; see
more on this in Section \ref{secinitvalue}.

\subsection{Pseudo-likelihood conditional on node degrees}

For networks with hub nodes or those with substantial degree variability
within communities, the block model can provide a poor fit, essentially
dividing the nodes into low-degree and high-degree groups. This has
been both observed empirically \cite{Karrer10} and supported by
theory \cite{Zhaoetal2012}.
The extension of the block model designed to cope with this situation,
the degree-corrected block model \cite{Karrer10}, has an extra degree
parameter to be estimated for every node, and writing out a pseudo-likelihood
that lends itself to an EM-type optimization is more complicated.
However, there is a simple alternative: consider the pseudo-likelihood
conditional on the observed node degrees. Whether these degrees are
similar or not will not then matter, and the fitted parameters will
reflect the underlying block structure rather than the similarities
in degrees.

The conditional pseudo-likelihood is again based on a simple observation:
\begin{longlist}[(C)]
\item[(C)] If random variables $X_{k}$ are independent Poisson
with means $\mu_{k}$, their distribution conditional on $\sum_{k}X_{k}$
is multinomial.
\end{longlist}
Applying this observation to the variables $(b_{i1},\ldots,b_{iK})$,
we have that their distribution, conditional on labels $c$ with $c_{i}=l$
and the node degree $d_{i}=\sum_{k}b_{ik}$, is multinomial with parameters
$(d_i; \theta_{l1}, \ldots, \theta_{lK})$, where $\theta_{lk}=\frac
{\lambda_{lk}}{\lambda_{l}}$. The conditional log
pseudo-likelihood (up to a constant)
is then given by
%
\begin{equation}
\label{cpl} \ell_{\CPL}\bigl(\pi,\Theta;\{\bb_{i}\}\bigr)=\sum
_{i=1}^{n}\log \Biggl(\sum
_{l=1}^{K}\pi_{l}\prod
_{k=1}^{K}\theta_{lk}^{b_{ik}}
\Biggr),
\end{equation}
and the parameters can be obtained by maximizing this function via
the EM algorithm for mixture models, as before. We again repeat the
EM for a fixed number of iterations, updating the initial partition
vector after the EM has converged. The algorithm is then the same
as that for unconditional pseudo-likelihood, with steps 2 and 3 replaced
by:
\begin{longlist}[{{$(3)$}}]
\item[{{$(2')$}}] Based on current estimates $\pih$ and $\{
\thetah_{lk}\}$,
let
\[
\pih_{il}=\prcpl(c_{i}=l | \bb_{i})=
\frac{\pih_{l}\prod_{m=1}^{K}\thetah_{lm}^{b_{im}}}{\sum_{k=1}^{K}\pih_{k}\prod_{m=1}^{K}\thetah_{km}^{b_{im}}}.
\]

\item[{{$(3)$}}] Given label probabilities, update parameter values
as follows:
\[
\pih_{l}=\frac{1}{n}\sum_{i=1}^{n}
\pih_{il},\qquad \thetah _{lk}=\frac{\sum_{i}\pih_{il}b_{ik}}{\sum_{i}\pih_{il}d_{i}}.
\]
\end{longlist}

\subsection{Initializing the partition vector}

\label{secinitvalue} We now turn to the question of how to initialize
the partition vector $e$. Note that the full likelihood,
pseudo-likelihoods $\ell_{\mathrm{PL}}$ and $\ell_{\mathrm{CPL}}$, and
other standard objective functions used for community detection such
as modularity \cite{Newman&Girvan2004} can all be multi-modal. The
numerical results in Section~\ref{secsimulations} suggest that the
initial value cannot be entirely arbitrary, but the results are not
too sensitive to it. We will quantify this further in Section \ref
{secsimulations};
here we describe the two options we use as initial values, both of
which are of independent interest as clustering algorithms for networks.

\subsubsection{Clustering based on 1- and 2-degrees}

One of the simplest possible ways to group nodes in a network is to
separate them by degree, say by one-dimensional $K$-means clustering
applied to the degrees as in \cite{Channarondetal2011}.
This only works for certain types of block models, identifiable from
their degree distributions, and in general $K$-means does
not deal well with data with many ties, which is the case with degrees.
Instead, we consider two-dimensional $K$-means clustering
on the pairs $(d_{i},d_{i}^{(2)})$, where $d_{i}^{(2)}$ is the number
of paths of length 2 from node $i$, which can be obtained by summing
the rows of $A^{2}$.

\subsubsection{Spectral clustering with perturbations}

A more sophisticated clustering scheme is based on spectral properties
of the adjacency matrix $A=\{A_{ij}\}$ or its graph Laplacian. Let
$D=\diag(d_{1},\ldots,d_{n})$ be diagonal matrix collecting node degrees.
A common approach\vspace*{1pt} is to look at the eigenvectors of the normalized
graph Laplacian $L=D^{-1/2}AD^{-1/2}$, choosing a small number, say
$r=K-1$, corresponding to
$r$ largest (in absolute value) eigenvalues, with the largest
eigenvalue omitted; see, for example, \cite{Shi00}. These vectors
provide an $r$-dimensional
representation for nodes of the graph, on which we can apply $K$-means
to find clusters; this is one of the versions of spectral clustering,
which was analyzed in the context of the block model in~\cite{Rohe2011}.

We found that this version of spectral clustering tends to do poorly
at community detection when applied to sparse graphs, say, with expected
degree $\lambda<5$. The $r$-dimensional representation seems to
collapse to a few points, likely due to the presence of many disconnected
components. We have found, however, that a simple modification performs
surprisingly well, even for values of $\lambda$ close to 1. The idea
is to connect
all disconnected components which belong to the same community by
adding artificial ``weak'' links. To be precise,
we ``regularize''
the adjacency matrix $A$ by adding $\alpha/p \times\lambda/n$
multiplied by the adjacency matrix of an Erdos--Renyi graph on $n$
nodes with edge probability $p$, where $\alpha$ is a constant. We
found that, empirically, $\alpha/p = 0.25$ works well for the range of
$n$ considered in our simulations, and that the results are
essentially the same for all $p > 0.1$ Thus we make the simplest and
computationally cheapest choice of $p=1$, adding a constant matrix
of small values, namely, $0.25 (\lambda/n) 1_n1_n^T$ where $1_n$ is
the all-ones $n$-vector, to the original adjacency matrix. The rest
of the steps, that is, forming the Laplacian, obtaining the spectral
representation and applying $K$-means, are performed on this regularized
version of $A$. We note that to obtain the spectral representation, one
only needs\vspace*{1pt} to know how the matrix acts on a given vector; since
$(A+0.25 (\lambda/n) 1_n1_n^T)x = Ax + 0.25(\lambda/n) (\sum_i x_i)
1_n$, the addition of the constant perturbation does not increase
computational complexity. We will refer to this algorithm as spectral clustering
with perturbations (SCP), since we perturb the network by adding
new, low-weight ``edges.''

%

\section{Consistency results}
\label{secconsist}


By consistency we mean consistency of node labels (to be defined
precisely below) under a block model
as the size of the
graph $n$ grows.
For the theoretical analysis, we only consider the case of $K = 2$
communities. We condition on the community labels
$\{c_i\}$, that is, we treat them as deterministic unknown parameters.
For simplicity, here we consider the case of balanced
communities, each having $m= n/2$ nodes. An extension to the
unbalanced case is provided in the supplementary material~\cite{Amietal}.
The assumption of balanced communities naturally leads us
to use the class prior estimates $\pih_1 = \pih_2 = 1/2$
in (\ref{eqcpl1stepiter}). We call this assumption (E)
(for equal class sizes):
\begin{longlist}[\mbox{(E)}]
\item[\mbox{(E)}] Assume each class contains $m= n/2$ nodes, and
set $\pih_1 = \pih_2 = 1/2$.
\end{longlist}
%
Without loss of generality,
we can take
$c_i = 1$ for $i \in\{1,2,\ldots,m\}$.

As an intermediate step in proving consistency for the block model
introduced in Section \ref{secintro}, we first prove the result for a
\emph{directed} block model. Recall that for the (undirected) block
model introduced earlier, one has
%
\begin{equation}
\label{equndirmod} \mbox{(undirected)}\quad A_{ij} \sim\ber(P_{c_i c_j})
\quad\mbox{and}\quad A_{ji} = A_{ij}\qquad \mbox{for } i \le
j.
\end{equation}
In the directed case, we assume that all the entries in the adjacency
matrix are drawn independently, that is,
%
\begin{equation}
\label{eqdirmod} \mbox{(directed)}\quad \Ad_{ij} \sim\ber(
\Pd_{c_i c_j}) \qquad\mbox{for all $i,j$}.
\end{equation}
We will use different symbols for the adjacency and edge-probability
matrices in the two cases. This is to avoid confusion when we need to
introduce a coupling between the two models. In both cases, we have
assumed that diagonal entries of the adjacency matrices are also drawn
randomly (i.e., we allow for self-loops as valid within-community
edges). This is convenient in the analysis with minor effect on the results.

The directed model is a natural extension of the block model when one
considers the pseudo-likelihood approach; in particular, it is the
model for which the pseudo-likelihood assumption of independence holds.
It is also a useful model of independent interest in many practical
situations, in which there is a natural direction to the link between
nodes, for example, in email, web, routing and some social networks.
The model can be traced back to the work of Holland and Leinhardt \cite
{HolLei81} and Wang and Wong \cite{wang1987} in which it has been
implicitly studied in the context of more general exponential families
of distributions for directed random graphs.

Our approach is to prove a consistency result for the directed model,
with an edge-probability matrix of the form
%
\begin{equation}
\label{eqedgeprobdir} \Pd= \frac{1}{m}\pmatrix{ a & b
\cr
b & a}.
\end{equation}
Note that the only additional restriction we are imposing is that
$\Pd$ has the same diagonal entries. Both $a$ and $b$ depend on $n$
and can in principle change with $n$ at different rates. This is a
slightly different parametrization from the more conventional $P_n =
\rho_n S$ \cite{Bickel&Chen2009}, where $S$ (and $\pi$) do not
depend on $n$,
and $\lambda_n = \rho_n \pi^T S \pi$. We use
this particular parametrization here because we only consider the case
$K=2$, and it makes our results more directly comparable to those
obtained in the physics literature, for example, \cite{Decelleetal2011}.

A coupling between the directed and the undirected model that we will
introduce allows us to carry the consistency result over to the
undirected model, with the edge-probability matrix
%
\begin{equation}
\label{eqedgeprobundir} P= \frac{2}{m}\pmatrix{ a & b
\cr
b & a}-
\frac{1}{m^2}\pmatrix{ a^2 & b^2
\cr
b^2 &
a^2}.
\end{equation}
Asymptotically, the two edge-probability matrices have comparable (to
first order)
expected degree and out-in-ratio (as defined by \cite
{Decelleetal2011}), under mild assumptions. The average degrees for
$\Pd$ and $P$ are $a + b$ and $2(a+b) -\frac{1}{m}(a^2 + b^2)$,
respectively. The latter is $\sim2(a+b)$ as long as $\frac1{2m}
\frac{a^2 + b^2}{a+b} \le\frac{a+b}{n}\to0$. The condition is
satisfied as soon as the average degree of the directed model has
sublinear growth: $a + b = o(n)$. The same holds for out-in-ratios.

For our analysis, we consider an E-step of the CPL algorithm. It starts
from some initial estimates $\ah$, $\bh$ and $\pih= (\pih_1,\pih
_2)$ of parameters $a$, $b$ and $\pi$, together with an initial
labeling $e$, and outputs the label estimates
%
\begin{equation}
\label{eqcpl1stepiter} \ch_i(e) = \arg\max_{k \in\{1,2\}}
\Biggl\{ \log\pih_k + \sum_{\ell= 1}^2
b_{i \ell}(e) \log\thetah_{k \ell}(e) \Biggr\},\qquad i \in[n],
\end{equation}
where $\thetah_{k\ell}$ are the elements of the matrix obtained by row
normalization of $\Lamh= [nR(e) \Ph]^T$. Here $R = R(e)$ is the
confusion matrix as defined in (\ref{eqRdef}), and $\Ph$ is given by
either (\ref{eqedgeprobdir}) or (\ref{eqedgeprobundir}),
depending on the model, with $a$ and $b$ replaced with their estimates
$\ah$ and $\bh$.

The key assumption of our analysis is that the initial labeling has a
certain overlap with the truth (we will show later that the amount of
overlap is not important). One situation where this might naturally
arise is survey data, when some small fraction of nodes has been
surveyed about their community membership. Another possibility is to
run some other crude algorithm first to obtain a preliminary result.
More formally, we
consider an initial labeling $e = (e_i) \in\{1,2\}^n$, which
is balanced (i.e., assigns equal number of nodes to each label) and
\emph{matches exactly $\gamma m$ labels in
community $1$}, for some $\gamma\in(0,1)$. We do not assume that
we know which labels are matched, or the value of $\gamma$. It is easy
to see that
this is equivalent to $e$ matching exactly $\gamma m$ labels in each
of the two communities. Assuming $\gamma m$ to be an integer, let
$\Ec^\gamma= \Ec^\gamma_n$ denote the collection
of such labelings,
%
\begin{equation}
\Ec^\gamma= \Ec^\gamma_n= \Biggl\{ e \in\{1,2
\}^n\dvtx \sum_{i=1}^m1_{\{
e_i = 1\}}
= \gamma m= \sum_{i=m+1}^n1_{\{e_i = 2\}}
\Biggr\}.
\end{equation}
%

Our goal is to obtain a uniform result guaranteeing the consistency of
CPL iteration (\ref{eqcpl1stepiter}) for any initial labeling in
$\Ec^\gamma$. In particular, this guarantees consistency for any initial
labeling of strength at least $\gamma$, even if it is obtained by an
algorithm operating on the same adjacency matrix used by\vspace*{1pt} CPL. As will
become clear in the course of the proof of Theorem \ref{thmcpldir},
although $\{\thetah_{k \ell}\}$ depend on $R(e)$ (which in turn
depends on $\gamma$) and $\Ph$, under the stated (idealized)
assumptions, we do not need to know their exact values in order to
implement rule (\ref{eqcpl1stepiter}). In particular, we do not
need to know $\gamma$. We can plug in any number in $(0,1)
\setminus\{\frac12\}$ for $\gamma$ and get the same estimates. Note
that the value of $\gamma= 1/2$ corresponds to ``no correlation''
between the true and the initial labeling, whereas $\gamma= 0$ and
$\gamma= 1$ both
correspond to perfect correlation (the labels are either all true or
all flipped).

Let us consider the directed case first. As our measure of performance
(i.e., the loss function), we take the following (directed-case)
mismatch ratio
%
\begin{equation}
\label{eqMisddef} \Misd_n(e):= \min_{\permute\in\{(1\,2),(2\,1)\}}
\frac{1}{n
} \sum_{i=1}^n1\bigl\{
\ch_i(e) \neq\phi(c_i)\bigr\},
\end{equation}
where
$\ch_i(e)$ are computed based on the directed adjacency matrix
$\Ad$, and $\{(1\,2),(2\,1)\}$ is the set of permutations of
$\{1,2\}$, with $\permute$ accounting for the
fact that the labels assigned to the communities are only determined
up to a permutation. The counterpart for the undirected case is denoted by
$M_n(e)$. Note that the notion of consistency based on
convergence of this quantity matches the ``weak'' consistency
discussed in \cite{Zhaoetal2012}, rather than the ``strong''
consistency used by \cite{Bickel&Chen2009}. Define
%
\begin{equation}
\tau^2_n= \frac{(a-b)^2}{a+b}
\end{equation}
and let $h(p) = -p\log p -(1-p)\log(1-p)$, $p \in[0,1]$ be the binary
entropy function. Let us also consider the collection of estimates
$(\ah,\bh)$ which have the same ordering as true parameters $(a,b)$,
\[
\Pc_{a,b} = \bigl\{ (\ah,\bh)\dvtx (\ah-\bh) (a-b) > 0 \bigr\}.
\]
Then, we have the following result.
%
\begin{theorem}[(Directed case)]\label{thmcpldir}
Assume \textup{(E)}, and let $\gamma\in(0,1) \setminus\{\frac12\}$. Let the
adjacency matrix $\Ad$ be generated according to the directed
model (\ref{eqdirmod}) with edge-probability matrix (\ref
{eqedgeprobdir}), and assume $a \neq b$. Then, there exists a
sequence $\{u_n\} \subset\reals_+$ such that
%
\begin{equation}
\label{equnineq} \log u_n+ \log\log u_n\ge\log \biggl(
\frac4e h(\gamma) \biggr) + \frac14(1-2\gamma)^2 \tau_n^2
\end{equation}
and
%
\begin{equation}
\pr \biggl[ \sup_{(\ah,\bh) \in\Pc_{a,b}} \sup_{e \in\Ec
^\gamma_n
}
\Misd_n(e) \ge\frac{4 h(\gamma)}{\log u_n} \biggr] \le\exp \bigl( {-n\bigl[h(
\gamma)-\kappa_\gam(n)\bigr] } \bigr),
\end{equation}
where $ \kappa_\gam(n):= \frac1n  [ \log (
\frac{n}{4 \pi\gam(1-\gam)} ) + \frac1{3n} ] = o(1)$.

In particular, if $\tau_n^2 \to\infty$, we have $u_n\to
\infty$ and the CPL estimate is uniformly consistent.
\end{theorem}

\begin{rem} We think of $\gamma$ as fixed, but it is possible to let
$\gamma= \gamma_n\to\frac12$, making the problem harder as $n
$ grows. We still get consistency as long as $(1-2\gamma_n)^2 \tau
_n^2 \to\infty$.
\end{rem}

\begin{rem}
In the balanced case, the CPL iteration has a simple intuitive
interpretation, as will become clear during the proof of
Theorem \ref{thmcpldir}. One starts with an initial assignment of
labels to nodes. Then, each node updates its label by taking a
majority vote among its neighbors. In the case where $ b = 0$, it is
intuitively clear that for $a$ large enough, this procedure
%
\begin{figure}

\includegraphics{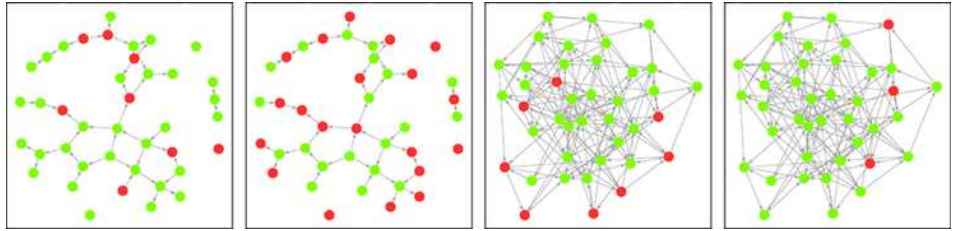}

\caption{The plots illustrate the interpretation of CPL iteration as
neighborhood majority voting, in the balanced case. Here $b=0$ and
only one community is shown. From left to right, we have the
initial labeling for a sparse graph $G_1$, the new labeling for
$G_1$ after one CPL iteration, the initial labeling for a dense
graph $G_2$, and the new labeling for $G_2$ after CPL
iteration. Nodes with red labels are ``infected,'' that is, their
community label is incorrect. For the sparse case, CPL iteration
spreads the infection, while for the dense case, it has the
opposite effect.}\label{figmajorvote}
\end{figure}
increases the number of correct labels relative to the initial
assignment. Figure \ref{figmajorvote} illustrates these ideas. In
the general case where $b \neq0$,
Theorem \ref{thmcpldir} states that $\tau_n^2$ is the key
parameter that needs to grow for the procedure to succeed.\looseness=1
\end{rem}

\begin{rem}
While the labels are of primary interest in community detection, one
may also be interested in consistency of the estimated parameters.
Under strong consistency in the sense of \cite{Bickel&Chen2009},
consistency of the natural plug-in estimates of the block model
parameters follows easily, but here we only show weak consistency of
the labels. However, in the directed model the pseudo-likelihood
function we defined
is in fact exactly the likelihood of $\bb_i$'s. Parameter estimates
(say $\ah$ and~$\bh$) obtained by the EM algorithm converge to a
local maximum of this function. As a consequence of
Theorem~\ref{thmcpldir}, these estimates are also consistent
(for $a$ and $b$). Since the likelihood is smooth with bounded
derivatives, one may be able to use standard arguments to show
that the estimated parameters are a unique local maximum in a
neighborhood of the truth, and even derive their asymptotic
normality along; see, for example, Theorem 6.2.1,
page 384 of \cite{Bickel&Doksum}. We do not pursue this direction
here.
\end{rem}

We now turn to the undirected case. Let
%
\begin{equation}
\label{eqdefagam} \agam= \gamma a+(1-\gamma) b.
\end{equation}
%
\begin{theorem}[(Undirected case)]\label{thmcplundir}
Assume \textup{(E)}, and let $\gamma\in(0,1) \setminus\{\frac12 \}$. Let
the adjacency matrix $A$ be generated according to the undirected
model (\ref{equndirmod}) with edge-probability matrix (\ref
{eqedgeprobundir}), and assume $a \neq b$.
In addition, assume
%
\begin{equation}
\label{eqagamassump} 2(1+\eps) \agam\le\eps(1-2\gamma) (a-b)
\end{equation}
for some $\eps\in(0,1)$. Then, there exist sequences $\{u_n\}, \{
v_n\} \subset\reals_+$ such that $\{u_n\}$ satisfies (\ref
{equnineq}), with $1-2\gamma$ replaced with $(1-\varepsilon)(1-2\gamma)$
and $\{v_n\}$ satisfies
\[
\log v_n+ \log\log v_n\ge\log
\biggl(\frac4e h(\gamma) \biggr) + \frac{\eps^2}{1+\eps/3} \agam
\]
and
%
\begin{eqnarray}
&&
\pr \biggl[ \sup_{(\ah,\bh) \in\Pc_{a,b}} \sup_{e \in\Ec
^\gamma_n
}
M_n(e) \ge4 h(\gamma) \biggl( \frac{1}{\log u_n} + \frac{2}{\log v_n}
\biggr) \biggr]\nonumber\\[-8pt]\\[-8pt]
&&\qquad \le3\exp \bigl( {-n\bigl[h(\gamma)-\kappa_\gam(n)
\bigr] } \bigr),\nonumber
\end{eqnarray}
where $\kappa_\gam(n) = o(1)$ is as defined in Theorem \ref{thmcpldir}.

In particular, if $\tau_n^2, \agam\to\infty$, we have
$u_n,v_n\to\infty$, and the CPL estimate is uniformly consistent.
\end{theorem}

The proofs of both theorems can be found in Section \ref{secproofs}.

\begin{rem} Condition (\ref{eqagamassump}) can be met
for a fixed $\eps\in(0,1)$ by choosing $\gamma$ sufficiently small
and an upper bound on $b/a$ in terms of $\gamma$. For example, for
$\eps= \frac12$ and $\gamma< \frac18$, we have (\ref
{eqagamassump}) if
\[
\frac{b}{a} \le\frac{1-8\gamma}{7-8\gamma}.
\]
\end{rem}

\begin{rem}
\label{weakerassum}
The parameter $\tau_n^2$ controlling consistency is the same as the
one reported in \cite{Decelleetal2011}
and \cite{Mosseletal2012}. There the concern is with recovering a
labeling which is positively correlated with the truth, and the
threshold of success is observed to be $\tau_n^2 \ge2$. A
similar lower bound was given in
\cite{Chaudhuri&Chung&Tsiatas2012} for spectral clustering.
Here, we are concerned with moving from a positively correlated
labeling to
one with an asymptotically vanishing mismatch ratio [i.e.,
$\Misd_n(e) = o_p(1)$], which is why we need $\tau_n^2 \to\infty$.
\end{rem}

\begin{rem}\label{remsuppAref}
These results can be extended to the
case of unbalanced communities. Such an extension is provided for
the directed block model in the supplementary material \cite{Amietal}.
There we consider the
model with two communities of sizes $n_1$ and $n_2$ (not necessarily
equal) and an edge-probability matrix
\[
\Pt= \frac{1}{n} \pmatrix{ a_1 & b
\cr
b & a_2},
\]
which relaxes our earlier assumption $a_1 = a_2$ in (\ref
{eqedgeprobdir}). The class of initial labelings is also enlarged to
include those that have $\gam_k$-overlap with community $k$, that is,
$\Ec^{\gam_1,\gam_2}:= \{ e\dvtx  \sum_{i} 1_{\{e_i = k, c_i = k\}} = \gam_k
n_k, k=1,2\}$, with $\gamma_1 \neq\gamma_2$. In this situation, one
needs more assumptions on the initial estimate $\Ph$ used in the CPL
iteration than in the balanced case. Supplementary material \cite{Amietal}
gives the details. While we do not discuss the undirected case
in this general setting, ideas used in the proof of Theorem
\ref{thmcplundir} can be used to carry the results from the directed to
the undirected case.
\end{rem}

\section{Numerical results}
\label{secsimulations}

Here we investigate the performance of both the unconditional and
conditional pseudo-likelihood algorithms on simulated networks, as
well as that of spectral clustering with perturbations. We
simulate two scenarios, one from the regular stochastic block model
and one from the degree-corrected block model, to assess the
performance in the presence of hub nodes. Throughout this section,
we fix $K=3$ and $\pi= (1/3, 1/3, 1/3)$. Conditional on the
labels, the edges are generated as independent Bernoulli variables
with probabilities proportional to $\theta_i \theta_j P_{ij}$. The
parameters $\theta_j$ are drawn independently from the distribution
of $\Theta$ with $\pr(\Theta= 0.2) = \rho$, $\pr(\Theta= 1) = 1-
\rho$. We do not enforce the identifiability scaling constraint on
$\theta$
at this point as it is absorbed into the scaling of the matrix $P$ in
(\ref{eqPscaling}) below. We consider two values of $\rho$:  $\rho= 0$,
which corresponds to the regular block model, and $\rho= 0.9$,
which corresponds to a network where 10\% of the nodes can be viewed
as hubs.

The matrix $P$ is constructed as follows. It is controlled by two
parameters: the ``out-in-ratio'' $\beta$ \cite{Decelleetal2011},
which we will vary from 0
to 0.2, and the weight vector~$w$, which determines the relative
degrees within communities. We consider two values of $w$: $w =
(1,1,1)$ (no information about communities is contained in node
degrees) and $w = (1, 5, 10)$ (degrees themselves provide relevant
information for clustering). If $\beta= 0$, we set $P^{(0)} =
\operatorname{diag}(w)$, a diagonal matrix. Otherwise, we set the diagonal of
$P^{(0)}$ to $\beta^{-1}w$ and set all off-diagonal elements to $1$.
We then fix the overall expected network degree $\lambda$, which is
the natural parameter to control \cite{Bickel&Chen2009} and which we
will vary from 1 to 15. Then we rescale $P^{(0)}$ to obtain this
expected degree, giving the final~$P$
%
\begin{equation}
\label{eqPscaling} P = \frac{\lambda}{(n-1) (\pi^T P^{(0)} \pi)(\ex\Theta)^2} P^{(0)}.
\end{equation}

To compare our results to the true labels, we will use normalized
mutual information (NMI). One can think of the confusion matrix $R$ as a
bivariate probability distribution, and of its row and column sums
$R_{i+}$ and $R_{+j}$ as the corresponding marginals. Then the
NMI is defined by \cite{Yao03} as $
\operatorname{NMI}(c,e) = - \sum_{i,j} R_{ij} \log\frac{R_{ij}}{R_{i+}
R_{+j}} (\sum_{i,j} R_{ij} \log R_{ij})^{-1}$,
and is always a number between 0 and~1 (perfect match).
It is useful to have a few benchmark values of NMI for reference: for
example, for large $n$, matching $50\%$, $70\%$ and $90\%$ of the
labels correspond to
values of NMI of approximately $0.12$, $0.26$ and $0.58$, respectively.

All figures show the performance of the following methods: $K$-means
clustering on 1- and 2-degrees (DC), spectral clustering (SC),
spectral clustering with perturbations (SCP), unconditional
pseudo-likelihood (UPL) initialized with either DC or SCP, and
conditional pseudo-likelihood (CPL), with the same two initial values
for labelings. The number of outer
iterations for UPL and CPL is set to $T=20$; $n$, $\lambda$, $\rho$
and the number of replications
$N$ are specified in the figures.

\begin{figure}

\includegraphics{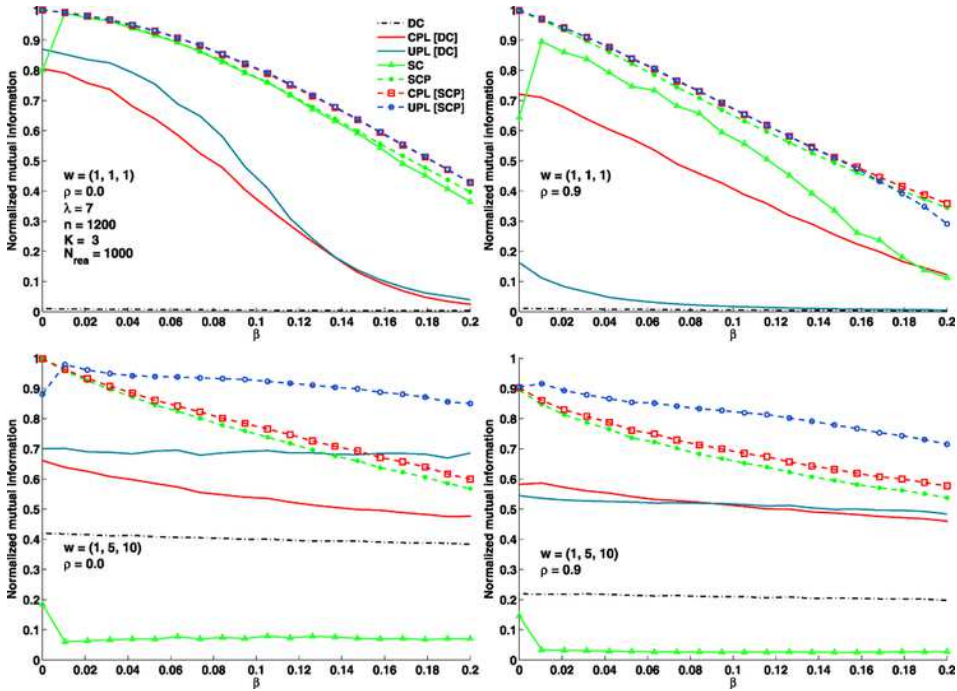}

\caption{The NMI between true and
estimated labels as a function of ``out-in-ratio'' $\beta$.}
\label{fignmibeta}
\end{figure}

\begin{figure}

\includegraphics{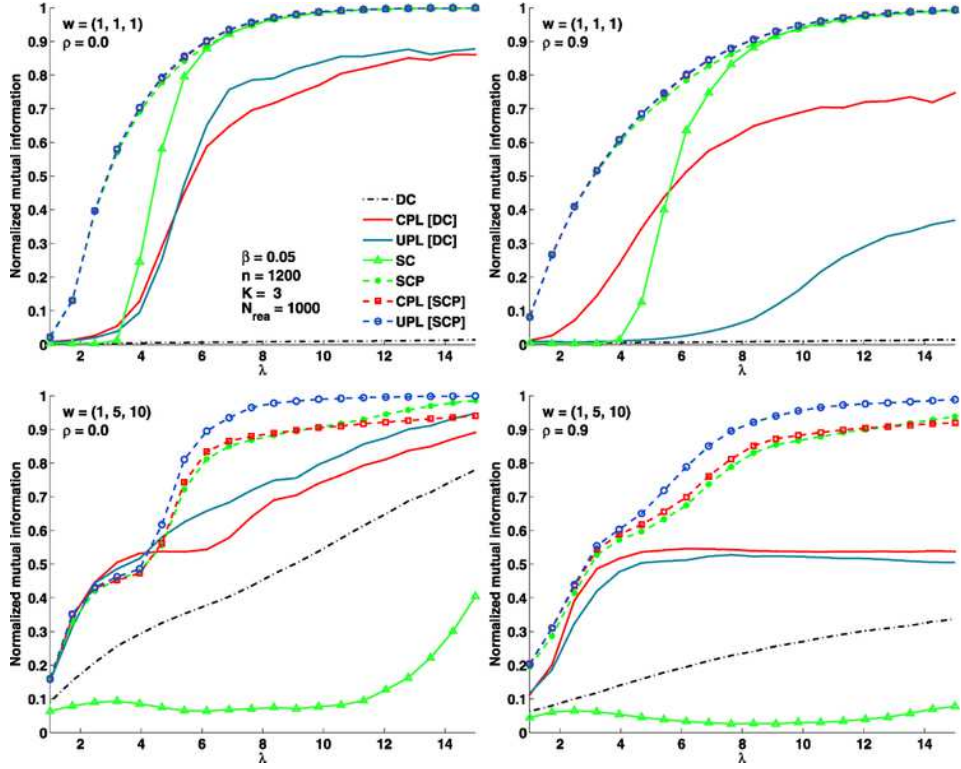}

\caption{The NMI between true and
estimated labels as a function of average expected degree $\lambda$.}
\label{fignmilam}
\end{figure}

Figures \ref{fignmibeta} and \ref{fignmilam} show results on
estimating the node labels with varying $\beta$ and~$\lambda$,
respectively. Generally, smaller $\beta$ and larger $\lambda$ make
the problem
easier, as we expect. In principle, degree-based clustering gives no
information about the labels
with uniform weights $w$, and only a moderate amount of information
with nonuniform weights, so it
serves as an example of a poor starting value for pseudo-likelihood.
Regular spectral clustering
performs well with uniform weights, but very poorly with nonuniform
weights; we conjecture that this is due to a limitation of $K$-means.
Spectral clustering with perturbation, on the other hand,
performs very well in all scenarios. Apart from being a useful
general method on its own, it also serves as an example of a good
starting value for pseudo-likelihood.

Figures \ref{fignmibeta} and \ref{fignmilam} show that
pseudo-likelihood achieves large gains over a poor starting value,
giving surprisingly good results even when starting from the
uninformative degree clustering in the case of $w = (1,1,1)$. One
exception is
unconditional pseudo-likelihood with $\rho=0.9$ and $w = (1,1,1)$,
which shows that
conditioning is necessary to accommodate variation in degrees when the
starting value is not very good. When
spectral clustering with perturbation is used as a starting value,
which is already very good, UPL
and CPL do not have much room to do better, although UPL still
provides a noticeable improvement, being overall the best method when
initialized with SCP. It appears that a good starting value
overcomes the limitations of the regular block model for networks with
hubs, effectively ruling out the competing solution which divides
nodes by degree.



\begin{figure}

\includegraphics{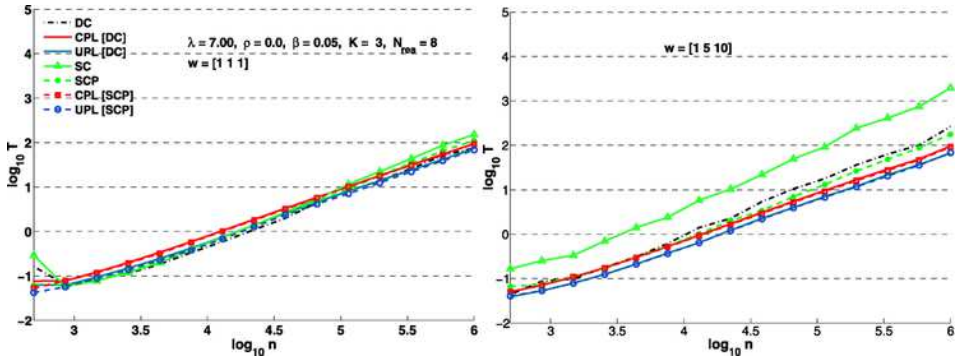}

\caption{The runtime in seconds
as a function of the number of nodes (log--log scale).}
\label{figtime}
\end{figure}

Finally, Figure \ref{figtime} shows run times for all the methods for
the case of the regular block model ($\rho= 0$) with different
community weights [$w = (1,1,1)$ and $w = (1,5,10)$].
The times shown for UPL and CPL do not include the
time to compute the initial value, which is shown
separately. For the case $w = (1,1,1)$, all methods take roughly the
same amount of time. For the case $w =
(1,5,10)$, spectral clustering (SC) takes considerably more time than the
rest. On the other hand, SCP takes nearly the same time as it takes
for $w =
(1,1,1)$, and it slightly outperforms DC for larger values of
$n$. This might be explained, in part, by the sparse matrix
multiplication required for DC, which is both time and memory-consuming for
large $n$. Generally, SCP provides an excellent starting value, with
low computational complexity in a variety of situations.

We have also done some brief comparisons with the belief propagation (BP)
method of \cite{Decelleetal2011}. Direct fair comparison is
difficult because of the
different platform for the belief propagation code and the different
way in which it handles initial values; generally, we found
that while the computing time of belief propagation scales with $n$ at
the same rate
as ours, BP is slower by a constant factor of about 10. In terms of
accuracy of community detection, in the examples we tried BP was
either similar to or a little worse than pseudo-likelihood.



\section{Example: A political blogs network}
\label{secexample}

This dataset on political blogs was compiled by Adamic and Glance \cite
{Adamic05} soon
after the 2004 U.S. presidential election. The nodes are blogs focused
on US politics, and the edges are hyperlinks between these blogs. Each
blog was manually labeled as liberal or conservative in
\cite{Adamic05}, and we treat these as true community labels. Following
\cite{Karrer10}, we ignore directions of the hyperlinks and analyze
the largest connected component of this network, which has 1222 nodes
and the average degree of 27. The distribution of degrees is highly
skewed to the right (the median degree is 13, and the maximum is
351).

\begin{figure}

\includegraphics{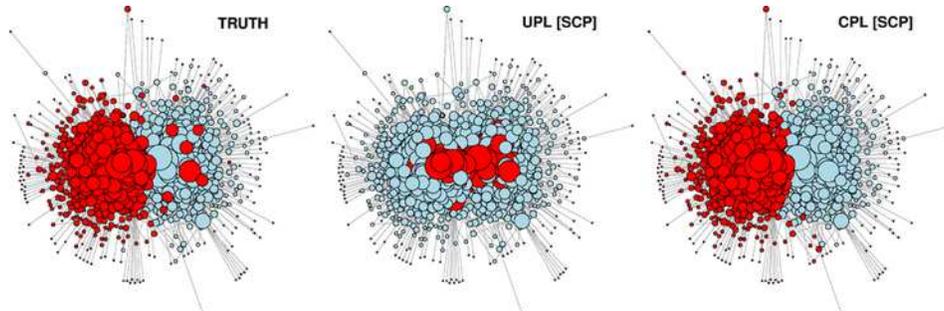}

\caption{Political blogs data: true labels and unconditional and conditional
pseudo-likelihoods (UPL and CPL) initialized with spectral
clustering with perturbations (SCP). Node size is proportional to log degree.}
\label{figblogs}
\end{figure}

The results in Figure \ref{figblogs} show that the conditional
pseudo-likelihood produces a result closest to the truth, as one would
expect in view of highly variable degrees. Its result is also very
close to those obtained by profile maximum likelihood for the
degree-corrected block model and by two different modularities
\cite{Karrer10,Zhaoetal2012}. Unconditional pseudo-likelihood, on the
other hand,
puts high-degree nodes in one group and low-degree nodes in the other.
This is very close to the block model solution \cite{Karrer10}. This
example confirms that the unconditional and conditional
pseudo-likelihood methods are correctly fitting the block model and
the degree-corrected block model, respectively.

\section{Proofs of consistency results}
\label{secproofs}
Due to symmetry, we can assume without loss of generality that $\gamma
\in(0,\frac12)$. Similarly, we can assume $a > b$. Then, for any
$(\ah,\bh) \in\Pc_{a,b}$ we have $\ah> \bh$. These will be our
standing assumptions throughout the proofs. To see that the
assumptions are not restrictive, one can check that the proof goes
through, without change, if $\gamma\in(\frac12,1)$ and $b > a$. For
the other two cases, namely, $\gamma\in(0,\frac12)$ and $ b > a$, or
$\gamma\in(\frac12,1)$ and $a > b$, the proof goes through by
switching the estimated labels when matching them with the true
labels. That is, we compare estimated community $1$ to true
community $2$ and vice versa. These can seen by
examining (\ref{eqcpl1simp}) and the discussion that follows.

\subsection{\texorpdfstring{Proof of Theorem \protect\ref{thmcpldir} (directed case)}
{Proof of Theorem 1 (directed case)}}
Let us introduce the following notation:
\begin{eqnarray*}
\Cc_\ell&=& \{i\dvtx c_i = \ell\},
\\
\Sc_k &=& \Sc_k(e) = \{i\dvtx e_i = k\},
\\
\Sc_{k \ell} &=& \Sc_{k \ell}(e) = \Sc_k \cap
\Cc_\ell
\end{eqnarray*}
for $k,\ell= 1,2$. As long as $e \in\Ec^\gamma$, we have $|\Cc
_\ell| =
|\Sc_k| = m$ for all $k,\ell= 1,2$ and
%
\begin{equation}
\label{eqSklcard} |\Sc_{11}| = |\Sc_{22}| = \gamma m,\qquad |
\Sc_{12}| = |\Sc_{21}| = (1-\gamma) m.
\end{equation}

Under the equal priors assumption (E), the CPL estimate (\ref
{eqcpl1stepiter}) simplifies to
\[
\ch_i(e) = \arg\max_{k \in\{1,2\}} \Biggl\{ \sum
_{m = 1}^2 \bd_{im}(e) \log
\thetah_{k m}(e) \Biggr\},
\]
where $\{\bd_{im} \}$ are obtained by block compression of the
directed adjacency matrix~$\Ad$.

Let us focus on $i \in\Cc_1$ from now on. Then $\ch_i(e) = 1$ if
%
\begin{equation}
\label{eqcpl1simp} \bd_{i1}(e) \log\frac{\thetah_{1 1}(e)}{\thetah_{2 1}(e)} +
\bd_{i2}(e) \log\frac{\thetah_{1 2}(e)}{\thetah_{2 2}(e)} > 0.
\end{equation}
For $e \in\Ec^\gamma$, we have $r_{k\ell}(e) = n^{-1} |\Sc_{k\ell}|$,
implying that
\[
R(e) = \frac12 \pmatrix{ \gamma& 1-\gamma
\cr
1-\gamma& \gamma},
\]
where $R(e)$ is
defined in (\ref{eqRdef}). It is then not hard to see that after row
normalization of $\Lamh= [nR(e) \Ph]^T$, we obtain
$\thetah_{11}(e) = \thetah_{22}(e) = \gamma\frac{\ah}{\ah+ \bh} +
(1-\gamma) \frac{\bh}{\ah+\bh}$, and $\thetah_{12}(e) = \thetah_{21}(e)
= \gamma\frac{\bh}{\ah+\bh} + (1-\gamma) \frac{\ah}{\ah+\bh}$.

Since by assumption $\ah> \bh$ and $\gamma\in(0,\frac12)$, it
follows that $\thetah_{11} < \thetah_{21}$. Then, (\ref
{eqcpl1simp}) is equivalent to $\bd_{i1}(e) - \bd_{i2}(e) < 0$.
Recalling that $\bd_{ik}(e) = \sum_{j=1}^m\Ad_{ij}1\{e_i = k\} =
\sum_{j \in\Sc_k} \Ad_{ij}$, we can write the condition as
\[
\xid_i\bigl(\sigma(e)\bigr) = \sum_{j=1}^n
\Ad_{ij} \sigma_j(e) < 0 \qquad\mbox{where }
\sigma_j(e) = \cases{ 1, &\quad $e_j = 1$,
\cr
-1, &\quad
$e_j = 2$,}
\]
and $\sigma(e) = (\sigma_1(e), \ldots, \sigma_{n}(e))$. Let
$\Sigma^\gamma=
\Sigma^\gamma_n$ be the set of all $\sigma(e)$ with $e \in\Ec
^\gamma$, that is,
\[
\Sigma^\gamma= \Sigma^\gamma_n= \Biggl\{ \sigma\in
\{-1,1\}^n\dvtx \sum_{j=1}^m1\{
\sigma_j = 1\} = \gamma m \Biggr\}.
\]

For $\ell= 1,2$, let $\Misd_{n,\ell}(e) = \frac1m\sum_{i \in
\Cc_\ell} 1\{ \ch_i(e) \neq c_i \}$ be the fraction of mismatches
over community $\ell$. Note that the overall mismatch is
%
\begin{equation}
\label{eqoverallMis} \Misd_n(e) = \tfrac12 \bigl[\Misd_{n,1}(e)
+ \Misd_{n,2}(e)\bigr].
\end{equation}
Since we are focusing on $i \in\Cc_1$, we are concerned with
$\Misd_{n,1}(e)$. In a slight abuse of notation, $\Misd_n(e)$ in
(\ref{eqoverallMis}) is in fact an upper bound on the mismatch ratio as
defined in (\ref{eqMisddef}), since here we are using
a particular permutation---the identity.

Let us define, for $\sigma\in\{-1,+1\}^n$ and $r \ge0$,
\[
\Nisd_{n,1}(\sigma;r) = \sum_{i=1}^m1
\bigl\{ \xid_i(\sigma) \ge -r \bigr\}.
\]
Then we have
\[
\sup_{e \in\Ec^\gamma} \Misd_{n,1}(e) \le\sup
_{\sigma\in
\Sigma^\gamma} \frac{\Nisd_{n,1}(\sigma;0)}{m},
\]
where the inequality is due to treating the ambiguous case
$\xid_{i}(\sigma) = 0$ as error. We now set out to bound this in
probability. Let us start with a tail bound on $\xid_i(\sigma)$ for
fixed $\sigma$ and $i$.

\begin{lem} \label{lemxitailbound}
For any $\sigma\in\Sigma^\gamma$ and $ t \in(0, 3(a+b)]$, we have
%
\begin{equation}
\label{eqxitailbound} \pr \bigl[ \xid_i(\sigma) \ge-(1-2\gamma)
(a-b) +t) \bigr] \le\exp \biggl( {-\frac{t^2}{4(a+b)}} \biggr).
\end{equation}
\end{lem}
\begin{pf}
We apply the classical Bernstein inequality for sums of independent
bounded random variables. Let $\alpha_{ij} = \ex[\Ad_{ij}]$. Note
that $|\Ad_{ij}\sigma_j - \ex[\Ad_{ij}\sigma_j]| \le\max(\alpha
_{ij},1-\alpha_{ij}) \le1$. For $i \in\Cc_1$, we have
\begin{eqnarray*}
\ex\xid_i(\sigma) &=& \sum_{j=1}^n
\alpha_{ij} \sigma_j = \sum_{j \in\Sc_{11}}
\frac{a}{m}(1) +\sum_{j \in\Sc_{22}}
\frac{b}{m}(-1) + \sum_{j \in\Sc_{21}}
\frac{a}{m}(-1) + \sum_{j \in\Sc_{12}}
\frac{b}{m}(1)
\\
&=& (a-b)\gamma+ (-a+b) (1-\gamma) = -(1-2\gamma) (a-b),
\end{eqnarray*}
where $\Sc_{k \ell}$ is defined based on labeling $e$ which
correspond to $\sigma$. In addition, since $\var(\Ad_{ij}) \le
\alpha_{ij}$, we have
\[
v= \sum_{j=1}^n\var(\Ad_{ij}
\sigma_j) \le \sum_{j \in\Cc_1}
\alpha_{ij} + \sum_{j \in\Cc_2}
\alpha_{ij} = m\frac{a}{m} + m\frac{b}{m} = a+b.
\]
Bernstein inequality implies
\[
\pr \bigl[ \xid_i(\sigma) \ge\ex\xid_i(\sigma) + t
\bigr] \le\exp \biggl( {- \frac{t^2}{2(v+ t/3)}} \biggr).
\]
Noting that for $t/3 \le(a+b)$, we have $2(v + t/3) \le4(a+b)$
completes the proof.
\end{pf}

We also need a tail bound on $\Nisd_{n,1}(\sigma;r)$. Let us define
%
\begin{equation}
\label{eqpirdef} p_i(r) = \pr \bigl[\xid_i(\sigma)
\ge-r \bigr],\qquad \ppb_1(r) = \frac1m\sum_{i=1}^mp_i(r).
\end{equation}
Note that these probabilities do not depend on the particular value of
$\sigma\in\Sigma^\gamma$, due to symmetry.
We have the following lemma.
%
\begin{lem}\label{lemNisdtailbound}
For $u > 1/e$,
%
\begin{equation}
\label{eqNistailbound} \pr \biggl[ \frac1m\Nisd_{n,1}(\sigma;r) \ge e u
\ppb _1(r) \biggr] \le\exp \bigl( {- e m\ppb_1(r) u \log
u} \bigr).
\end{equation}
\end{lem}
\begin{pf}
Follows from Lemma \ref{lempoitail} in the \hyperref[app]{Appendix},
by noting that\break $ \{ 1\{\xid_i(\sigma) \ge-r\}  \}
_{i=1}^m$ are independent Bernoulli random variables.
\end{pf}

Now we apply Lemma \ref{lemxitailbound} with $t = (1-2\gamma)(a-b)
\le3(a+b)$. Note that $\frac{a-b}{a+b} \le1 \le\frac3{1-2\gamma}$,
for $\gamma\in(0,\frac12)$. Noting that the RHS of (\ref
{eqxitailbound}) does not depend on $i$, and using (\ref
{eqpirdef}), we get
\[
\ppb_1(0) \le\exp \biggl\{ {-\frac14(1-2\gamma)^2
\frac
{(a-b)^2}{a+b}} \biggr\}.
\]

The cardinality of the set $\Sigma^\gamma$ is ${m\choose\gamma m}^2
\le
(e^{m[h(\gamma) + \kappa_\gam(2 m)]})^2$ where $h(\cdot)$ is the
binary entropy function, and $\kappa_\gam(2 m) = \kappa_\gam(n)$ is
as defined in the statement of the theorem. (See Lemma 6 %
in the supplementary material \cite{Amietal} for a proof.) Applying
Lemma \ref{lemNisdtailbound} with $u = u_n$ and the union bound,
we obtain
\begin{eqnarray*}
&&
\pr \biggl[ \sup_{\sigma\in\Sigma^\gamma} \frac1m\Nisd _{n,1}(\sigma;0)
\ge e u_n\ppb_1(0) \biggr]\\
&&\qquad \le \exp \bigl\{ {m \bigl[
2h(\gamma) -e \ppb_1(0) u_n\log u_n+ 2\kappa
_\gam(n) \bigr]} \bigr\}.
\end{eqnarray*}
Pick $u_n$ such that
\[
u_n\log u_n= \frac{4h(\gamma)}{e \ppb_1(0)}.
\]
It follows, using $m= n/2$, that
\[
\pr \biggl[ \sup_{\sigma\in\Sigma^\gamma} \frac1m\Nisd _{n,1}(\sigma;0)
\ge \frac{4 h(\gamma)}{\log u_n} \biggr] \le\exp\bigl\{ -\bigl[h(\gamma )-
\kappa_\gam(n)\bigr] n\bigr\}.
\]
By symmetry the same bound holds for $\sup_{\sigma}\frac1m
\Nisd_{n,2}(\sigma;0)$. It follows from (\ref{eqoverallMis}) that
the same holds for $\sup_{e} M_n(e)$. This completes the proof
of Theorem \ref{thmcpldir}.

\subsection{\texorpdfstring{Proof of Theorem \protect\ref{thmcplundir} (undirected case)}
{Proof of Theorem 2 (undirected case)}}
Recall that $A$ and $\Ad$ are the adjacency matrices of the undirected
and directed cases, respectively. Let us define $\xi _i(\sigma)$,
$M_{n,\ell}(e)$, $N_{n,\ell}(\sigma,r)$ as we did in the directed case,
but based on $A$ instead of $\Ad$. For example, $\xi_i(\sigma) =
\sum_{j=1}^nA_{ij} \sigma_j$.

Our approach is to introduce a \emph{deterministic coupling} between
$A$ and $\Ad$, which allows us to carry over the results of the
directed case. Let
%
\begin{equation}
\label{eqcoupleAAt}
A= T(\Ad),\qquad
\bigl[T(\Ad)\bigr]_{ij} = \cases{ 0, &
\quad $\Ad_{ij} = \Ad_{ji} = 0$,
\cr
1, &\quad otherwise.}
\end{equation}
In other words, the graph of $A$ is obtained from that of $\Ad$ by
removing directions. Note that
\[
P_{kl} = \pr(A_{ij} = 1) = 1-\pr(\Ad_{ij} = 0)\pr(
\Ad_{ji} = 0) = 2\Pd_{kl} - \Pd_{kl}^2,
\]
which matches the relation between (\ref{eqedgeprobdir}) and (\ref
{eqedgeprobundir}). From (\ref{eqcoupleAAt}), we also note that
%
\begin{equation}
\label{eqcoupleineq} A_{ij} \ge\Ad_{ij}\qquad\mbox{for all
$i,j$}.
\end{equation}

Let us now upper-bound $\xi_i(\sigma)$ in terms of $\xid_i(\sigma
)$. Based on (\ref{eqcoupleineq}), only those $\sigma_j$ that are
equal to $1$ contribute to the upper bound. More precisely, let $D
_{ij} = A_{ij} - \Ad_{ij} \ge0$, and take $i \in\Cc_1$ from now
on. Then
%
\begin{eqnarray}
\label{eqxixitupbound} \xi_i(\sigma) - \xid_i(\sigma)
&=& \sum_{j \in\Sc_1} D_{ij}
\sigma_j + \sum_{j \in\Sc_2} D_{ij}
\sigma_j
\nonumber
\\
&=& \sum_{j \in\Sc_1} D_{ij} - \sum
_{j \in\Sc_2} D_{ij}
\\
&\le& \sum_{j \in\Sc_1} D_{ij}.\nonumber
\end{eqnarray}
We further notice that $D_{ij} \le\Ad_{ij} + \Ad_{ji}$. To
simplify notation, let us define
%
\begin{equation}
\label{eqdefagamDis} \Ais(\sigma) = \sum_{j \in\Sc_1}
\Ad_{ij},\qquad \Asi(\sigma) = \sum_{j \in\Sc_1}
\Ad_{ji},
\end{equation}
where the dependence on $\sigma$ is due to $\Sc_1$ being derived from
$\sigma$ [recall that $\Sc_1 = \Sc_1(\sigma) = \{j\dvtx  \sigma_j =
1\}$]. Thus we have shown
%
\begin{equation}
\label{eqxiaxitineq} \xi_i(\sigma) \le\xid_i(\sigma) +
\Ais(\sigma) + \Asi(\sigma).
\end{equation}
Recall from definition (\ref{eqdefagam}) that
$
\agam= \gamma a+(1-\gamma) b.
$

%
\begin{lem}\label{lemAistailbound}
Fix $\eps> 0$. For $i \in\Cc_1$, we have
\[
\pr \bigl[ \Ais(\sigma) > (1+\eps) \agam \bigr] = \pr \bigl[ \Asi(\sigma) > (1+
\eps) \agam \bigr] \le \exp \biggl\{ {-\frac{\eps^2}{1+\eps/3}} \agam \biggr\}.
\]
%
\end{lem}
\begin{pf}
The equality of the two probabilities follows by symmetry. Let us prove
the bound for $\Ais(\sigma)$. We apply Bernstein inequality. Note that
\begin{eqnarray*}
\mu&=&\ex \biggl[\sum_{j\in\Sc_1} \Ad_{ij}
\biggr] = \sum_{j \in\Sc_{11}} \ex[\Ad_{ij}]+ \sum
_{j \in\Sc_{12}} \ex [\Ad_{ij}]
\\
&=& \sum_{j \in\Sc_{11}} \frac{a}{m} + \sum
_{j \in\Sc_{12}} \frac{b}{m} = a\gamma+ b(1-\gamma) = \agam.
\end{eqnarray*}
Since $\sum_{j \in\Sc_1} \var(\Ad_{ij}) \le\mu$, we obtain
\[
\pr \biggl[ \sum_{j \in\Sc_1} \Ad_{ij}\ge\mu+ t
\biggr] \le\exp \biggl( {-\frac{t^2}{2(\mu+ t/3)}} \biggr).
\]
Setting $t = \eps\mu$ completes the proof.
\end{pf}

From (\ref{eqxiaxitineq}), it follows that
\[
\xi_i(\sigma) \ge0 \quad\implies\quad \bigl(\xid_i(\sigma) \ge-r
\bigr) \vee \bigl(\Ais(\sigma) \ge r/2\bigr) \vee \bigl(\Asi(\sigma) \ge r/2
\bigr),
\]
which $\vee$ is the logical OR. This can be seen (as usual) by noting
that if the RHS does not hold, then $\xid_i(\sigma) + \Ais(\sigma)
+ \Asi(\sigma) < 0$, implying $\xi_i(\sigma) < 0$. Translating to
indicator functions,
\[
1\bigl\{ \xi_i(\sigma) \ge0 \bigr\} \le1\bigl\{
\xid_i(\sigma) \ge-r \bigr\} + 1\bigl\{ \Ais(\sigma) \ge r/2 \bigr\}
+ 1\bigl\{ \Asi(\sigma) \ge r/2 \bigr\}.
\]
Averaging over $i \in\Cc_1$ (i.e., applying $m^{-1} \sum_{i=1}^m$), we get
%
\begin{equation}
\label{eqNQineq}\quad \frac{1}{m}N_{n,1}(\sigma;0) \le
\frac{1}{m}\Nisd_{n,1}(\sigma;r) + \frac{1}{m}
\Qu_{n,1*}(\sigma;r/2) + \frac{1}{m}\Qu_{n,*1}(
\sigma;r/2),
\end{equation}
where $\Qu_{n,1*}(\sigma;t) = \sum_{i=1}^m1\{ \Ad_{i*}(\sigma)
\ge t\}$, and similarly for $\Qu_{n,*1}(\sigma;t)$. Note that
$\Qu_{n,1*}(\sigma;t)$ and $ \Qu_{n,*1}(\sigma;t)$, while not
independent, have the same distribution by symmetry, so we can focus
on bounding one of them. The key is that each one is a sum of i.i.d.
terms, for example, $\{\Ad_{i*}\}_{i=1}^m$.

We have a bound on $m^{-1} \Nisd_{n,1}(\sigma;r)$ from
Lemma \ref{lemNisdtailbound}. We can get similar bounds on the $\Qu
$-terms. To start, let
%
\begin{equation}
q_i(r) = \pr \bigl[\Ais(\sigma) \ge r/2 \bigr],\qquad
\qqb_1(r) = \frac1m\sum_{i=1}^mq_i(r),
\end{equation}
similar to (\ref{eqpirdef}), and note that these quantities too
are independent of the particular choice of $\sigma\in\Sigma^\gamma$.

\begin{lem}\label{lemQtailbound}
For $u > 1/e$,
%
\begin{equation}
\label{eqQtailbound} \pr \biggl[ \frac1m\Qu_{n,1*}(\sigma;r/2) \ge e u
\qqb _1(r) \biggr] \le\exp \bigl( {- e m\qqb_1(r) u \log
u} \bigr).
\end{equation}
\end{lem}
\begin{pf}
Follows from Lemma \ref{lempoitail} in the \hyperref[app]{Appendix},
by noting that\break $ \{ 1\{ \Ad_{i*}(\sigma) \ge r/2\}  \}
_{i=1}^m$ is an independent sequence of Bernoulli variables.
\end{pf}

The same bound holds for $\frac{1}{m}\Qu_{n,*1}(\sigma;r /2)$. Recall
the definition of $\ppb_1(r)$ from~(\ref{eqpirdef}). Using
(\ref{eqNQineq}) and Lemmas \ref{lemNisdtailbound} and
\ref{lemQtailbound}, we get
\begin{eqnarray*}
&&\pr \biggl[ \sup_{\sigma\in\Sigma^\gamma} \frac1mN_{n,1}(\sigma;0)
\ge e \bigl[u_n\ppb_1(r) + 2 v_nq_1(r)
\bigr] \biggr]
\\
&&\qquad \le\pr \biggl[ \sup_{\sigma\in\Sigma^\gamma} \frac1m
\Nisd_{n,1}(\sigma;r) \ge e u_n\ppb_1(r) \biggr]\\
&&\qquad\quad{}+ 2 \pr \biggl[ \sup_{\sigma\in\Sigma^\gamma} \frac1m\Qu_{n,1*}(
\sigma;r/2) \ge e v_n\qqb_1(r) \biggr]
\\
&&\qquad \le \exp \bigl\{ {m \bigl[ 2h(\gamma) -e \ppb_1(r)
u_n\log u_n+ 2\kappa _\gam(n) \bigr]} \bigr\}\\
&&\qquad\quad{}+ 
2\exp \bigl\{ {m \bigl[ 2h(\gamma) -e \qqb_1(r)
v_n\log v_n+ 2\kappa _\gam(n) \bigr]} \bigr\}
\end{eqnarray*}
as long as $u_n,v_n> 1/e$. Now, take $r/2 = (1+\eps)\agam$, so
that Lemma \ref{lemAistailbound} implies
\[
\qqb_1(r) \le\exp \biggl\{ {-\frac{\eps^2}{1+\eps/3}} \agam \biggr\}.
\]
Now, in Lemma \ref{lemxitailbound}, take $t = (1-2\gamma)(a-b) -
2(1+\eps) \agam$. Note that the assumption
\[
2(1+\eps)\agam\le\eps(1-2\gamma) (a-b)
\]
implies $t \ge(1-\eps)(1-2\gamma)(a-b) > 0$. In addition $t \le
(1-2\gamma)(a-b) \le3(a+b)$ as before. Thus, the chosen $t$ is valid
for Lemma \ref{lemxitailbound}. Furthermore, $-(1-2\gamma)(a-b) +
t = -r$. Hence, the lemma implies
\[
\ppb_1(r) \le\exp \biggl\{ {-\frac14\bigl[(1-\eps) (1-2\gamma)
\bigr]^2\frac
{(a-b)^2}{a+b}} \biggr\}.
\]
Pick $u_n$ and $v_n$ such that
\[
u_n\log u_n= \frac{4h(\gamma)}{e \ppb_1(r)},\qquad v_n\log
v_n= \frac{4h(\gamma)}{e \qqb_1(r)}.
\]
The rest of the argument follows as in the directed case. This
completes the proof of Theorem \ref{thmcplundir}.

\section{Discussion}
\label{secdiscuss}



The proposed pseudo-likelihood algorithms provide
fast and accurate community detection for a range of settings,
including large and sparse networks, contributing to the long history
of empirical success of pseudo-likelihood
approximations in statistics. For the theoretical analysis, we did not
focus on the convergence
properties of the algorithms, since standard EM theory guarantees
convergence to a local maximum as long as the underlying Poisson or
multinomial mixture is identifiable. The consistency of a single
iteration of the algorithm was established for an initial value that
is better than purely arbitrary, as long as, roughly speaking, the
graph degree grows, and there are two balanced communities with equal
expected degrees. The theory shows that this local
maximum is consistent, and unique in a neighborhood of the truth, so
in fact there is no need to assume that EM has converged to the global
maximum, an assumption which is usually made in analyzing EM-based
estimates. The theoretical analysis can be extended to the general
two-community model with possibly unbalanced communities, as detailed
in the supplementary material \cite{Amietal}. Extending our argument to
more than two communities
also seems possible, but that would require extremely meticulous
tracking of a large number of terms which we did not pursue.

We conjecture that additional results may be obtained under weaker
assumptions if one focuses simply on estimating the parameters of the
block model rather than consistency of the labels, just like one can
obtain results
for a labeling correlated with the truth (instead of
consistent) under weaker assumptions discussed in Remark
\ref{weakerassum}. For example, in a very recent paper
\cite{Chatterjee2012}, results are obtained under very
weak assumptions for the mean squared error of estimating the block
model parameter matrix $P$ (which in itself does not guarantee
consistency of the labels). While the primary interest in
community detection is estimating the labels rather than the
parameters, we plan to investigate this further to see if and how our
conditions can be relaxed.

While in theory any ``reasonable'' initial value guarantees
convergence, in practice the choice of initial value is still
important, and we have investigated a
number of options empirically. Spectral clustering with perturbations,
which we introduced primarily
as a method to initialize pseudo-likelihood, deserves more study, both
empirically (e.g., investigating the optimal choice of the
tuning parameter), and theoretically. This is also a topic for future
work.

\begin{appendix}\label{app}
\section*{Appendix: Poisson-type tail bound}
Here is a lemma which we used quite often in proving consistency
results in Section \ref{secproofs}.
%
\begin{lem}\label{lempoitail}
Consider $X_1,X_2,\ldots,X_m$ to be independent Bernoulli variables
with $\ex[X_i] = p_i$. Let $S_m= \sum_{i=1}^mX_i$, $\mu=
\ex[S_m] = \sum_{i=1}^mp_i$ and $\mub= m^{-1} \mu$.
Then, for any $u > 1/e$, we have
\[
\pr \biggl( \frac1mS_m> e u \mub \biggr) \le\exp( - e m\mub u\log
u).
\]
\end{lem}

\begin{pf}
We apply a direct Chernoff bound. Let $S^*_m\sim\operatorname
{Bin}(m, \mub)$. Then, by a result of Hoeffding \cite{Hoe56} (also
see \cite{Gle75}), $\ex g(S_m) \le\ex g(S^*_m)$ for any convex
function $g\dvtx  \reals\to\reals$. Letting $g(x) = e^{\beta x}$, we
obtain for $\beta> 0$,
\begin{eqnarray*}
\pr( S_m> t) &\le& e^{-\beta t} \ex\bigl(e^{\beta S^*_m}\bigr)
= e^{-\beta t} \bigl( 1 + \mub\bigl(e^t - 1\bigr)
\bigr)^m
\\
&\le& e^{-\beta t} \exp \bigl\{ m \mub\bigl(e^t-1\bigr) \bigr\},
\end{eqnarray*}
where we have used $(1+x)^m \le\exp(mx)$. 
The RHS is the Chernoff bound for a Poisson random variable with mean
$\mu= \sum_i p_i$, and can be optimized to yield
\[
\pr( S_m> t) \le\frac{e^{-\mu}(e\mu)^t}{t^t} \qquad\mbox{for } t > \mu.
\]
Letting $t = eu\mu$ for $u > 1/e$ and noting that $e^{-\mu} \le1$,
we get $\pr(S_m> e u\mu) \le(1/u)^{e u\mu}$ which is the
desired bound.
\end{pf}
\end{appendix}

\section*{Acknowledgment}

We would like to thank Roman Vershynin (Mathematics, University of
Michigan) for highly illuminating discussions.

\begin{supplement}
\stitle{Extension to unbalanced communities}
\slink[doi]{10.1214/13-AOS1138SUPP} 
\sdatatype{.pdf}
\sfilename{aos1138\_supp.pdf}
\sdescription{This supplement contains an extension of
Theorem \ref{thmcpldir} to the case of unbalanced communities.}
\end{supplement}


\printaddresses


\begin{thebibliography}{39}

\bibitem{Adamic05}
\begin{bincollection}[auto:STB|2013/06/05|13:45:01]
\bauthor{\bsnm{Adamic},~\bfnm{L.~A.}\binits{L.~A.}} \AND
\bauthor{\bsnm{Glance},~\bfnm{N.}\binits{N.}}
(\byear{2005}).
\btitle{The political blogosphere and the 2004 US election}.
In \bbooktitle{Proceedings of the WWW-2005 Workshop on the Weblogging
Ecosystem}.
\bpublisher{ACM}, \blocation{New York}.
\bptok{imsref}%
\end{bincollection}
\endbibitem

\bibitem{Airoldi2008}
\begin{barticle}[auto:STB|2013/06/05|13:45:01]
\bauthor{\bsnm{Airoldi},~\bfnm{E.~M.}\binits{E.~M.}},
\bauthor{\bsnm{Blei},~\bfnm{D.~M.}\binits{D.~M.}},
\bauthor{\bsnm{Fienberg},~\bfnm{S.~E.}\binits{S.~E.}} \AND
\bauthor{\bsnm{Xing},~\bfnm{E.~P.}\binits{E.~P.}}
(\byear{2008}).
\btitle{Mixed membership stochastic blockmodels}.
\bjournal{J. Mach. Learn. Res.}
\bvolume{9}
\bpages{1981--2014}.
\bptok{imsref}%
\end{barticle}
\endbibitem

\bibitem{Amietal}
\begin{bmisc}[auto:STB|2013/06/05|13:45:01]
\bauthor{\bsnm{Amini},~\bfnm{A.~A.}\binits{A.~A.}},
\bauthor{\bsnm{Chen},~\bfnm{A.}\binits{A.}},
\bauthor{\bsnm{Bickel},~\bfnm{P.~J.}\binits{P.~J.}} \AND
\bauthor{\bsnm{Levina},~\bfnm{E.}\binits{E.}}
(\byear{2013}).
\bhowpublished{Supplement to ``Pseudo-likelihood methods for community
detection in large sparse networks.'' DOI:\doiurl{10.1214/13-AOS1138SUPP}}.
\bptok{imsref}%
\end{bmisc}
\endbibitem

\bibitem{Hall&Karrer&Newman2011}
\begin{barticle}[auto:STB|2013/06/05|13:45:01]
\bauthor{\bsnm{Ball},~\bfnm{B.}\binits{B.}},
\bauthor{\bsnm{Karrer},~\bfnm{B.}\binits{B.}} \AND
\bauthor{\bsnm{Newman},~\bfnm{M.~E.~J.}\binits{M.~E.~J.}}
(\byear{2011}).
\btitle{An efficient and principled method for detecting communities in
networks}.
\bjournal{Phys. Rev. E (3)}
\bvolume{34}
\bpages{036103}.
\bptok{imsref}%
\end{barticle}
\endbibitem

\bibitem{Besag74}
\begin{barticle}[mr]
\bauthor{\bsnm{Besag},~\bfnm{Julian}\binits{J.}}
(\byear{1974}).
\btitle{Spatial interaction and the statistical analysis of lattice systems}.
\bjournal{J. R. Stat. Soc. Ser. B Stat. Methodol.}
\bvolume{36}
\bpages{192--236}.
\bid{issn={0035-9246}, mr={0373208}}
\bptnote{check related}%
\bptok{imsref}%
\end{barticle}
\endbibitem

\bibitem{Bickel&Chen2009}
\begin{barticle}[auto:STB|2013/06/05|13:45:01]
\bauthor{\bsnm{Bickel},~\bfnm{P.~J.}\binits{P.~J.}} \AND
\bauthor{\bsnm{Chen},~\bfnm{A.}\binits{A.}}
(\byear{2009}).
\btitle{A nonparametric view of network models and Newman--{G}irvan and other
modularities}.
\bjournal{Proc. Natl. Acad. Sci. USA}
\bvolume{106}
\bpages{21068--21073}.
\bptok{imsref}%
\end{barticle}
\endbibitem

\bibitem{Bickel&Chen&Levina2011}
\begin{barticle}[mr]
\bauthor{\bsnm{Bickel},~\bfnm{Peter~J.}\binits{P.~J.}},
\bauthor{\bsnm{Chen},~\bfnm{Aiyou}\binits{A.}} \AND
\bauthor{\bsnm{Levina},~\bfnm{Elizaveta}\binits{E.}}
(\byear{2011}).
\btitle{The method of moments and degree distributions for network models}.
\bjournal{Ann. Statist.}
\bvolume{39}
\bpages{2280--2301}.
\bid{doi={10.1214/11-AOS904}, issn={0090-5364}, mr={2906868}}
\bptok{imsref}%
\end{barticle}
\endbibitem

\bibitem{Bickel&Choi&etal2012}
\begin{bmisc}[auto:STB|2013/06/05|13:45:01]
\bauthor{\bsnm{Bickel},~\bfnm{P.~J.}\binits{P.~J.}},
\bauthor{\bsnm{Choi},~\bfnm{D.}\binits{D.}},
\bauthor{\bsnm{Chang},~\bfnm{X.}\binits{X.}} \AND
\bauthor{\bsnm{Zhang},~\bfnm{H.}\binits{H.}}
(\byear{2012}).
\bhowpublished{Asymptotic normality of maximum likelihood and its variational
approximation for stochastic blockmodels. Available at
\arxivurl{arXiv:1207.0865}}.
\bptok{imsref}%
\end{bmisc}
\endbibitem

\bibitem{Bickel&Doksum}
\begin{bbook}[mr]
\bauthor{\bsnm{Bickel},~\bfnm{Peter~J.}\binits{P.~J.}} \AND
\bauthor{\bsnm{Doksum},~\bfnm{Kjell~A.}\binits{K.~A.}}
(\byear{2007}).
\btitle{Mathematical Statistics: Basic Ideas and Selected Topics},
\bedition{2nd} ed.
\bpublisher{Prentice Hall}, \blocation{New York}.
\bptok{imsref}%
\end{bbook}
\endbibitem

\bibitem{Celisseetal2011}
\begin{barticle}[mr]
\bauthor{\bsnm{Celisse},~\bfnm{Alain}\binits{A.}},
\bauthor{\bsnm{Daudin},~\bfnm{Jean-Jacques}\binits{J.-J.}} \AND
\bauthor{\bsnm{Pierre},~\bfnm{Laurent}\binits{L.}}
(\byear{2012}).
\btitle{Consistency of maximum-likelihood and variational estimators in the
stochastic block model}.
\bjournal{Electron. J. Stat.}
\bvolume{6}
\bpages{1847--1899}.
\bid{doi={10.1214/12-EJS729}, issn={1935-7524}, mr={2988467}}
\bptok{imsref}%
\end{barticle}
\endbibitem

\bibitem{Channarondetal2011}
\begin{bmisc}[auto:STB|2013/06/05|13:45:01]
\bauthor{\bsnm{Channarond},~\bfnm{A.}\binits{A.}},
\bauthor{\bsnm{Daudin},~\bfnm{J.~J.}\binits{J.~J.}} \AND
\bauthor{\bsnm{Robin},~\bfnm{S.}\binits{S.}}
(\byear{2011}).
\bhowpublished{Classification and estimation in the stochastic block model
based on the empirical degrees. Available at \arxivurl{arXiv:1110.6517}}.
\bptok{imsref}%
\end{bmisc}
\endbibitem

\bibitem{Chatterjee2012}
\begin{bmisc}[auto:STB|2013/06/05|13:45:01]
\bauthor{\bsnm{Chatterjee},~\bfnm{Sourav}\binits{S.}}
(\byear{2012}).
\bhowpublished{Matrix estimation by universal singular value thresholding.
Available at \arxivurl{arXiv:1212.1247}}.
\bptok{imsref}%
\end{bmisc}
\endbibitem

\bibitem{Chaudhuri&Chung&Tsiatas2012}
\begin{barticle}[auto:STB|2013/06/05|13:45:01]
\bauthor{\bsnm{Chaudhuri},~\bfnm{Kamalika}\binits{K.}},
\bauthor{\bsnm{Chung},~\bfnm{Fan}\binits{F.}} \AND
\bauthor{\bsnm{Tsiatas},~\bfnm{Alexander}\binits{A.}}
(\byear{2012}).
\btitle{Spectral clustering of graphs with general degrees in the extended
planted partition model}.
\bjournal{JMLR Workshop and Conference Proceedings}
\bvolume{23}
\bpages{35.1--35.23}.
\bptok{imsref}%
\end{barticle}
\endbibitem

\bibitem{Decelleetal2011}
\begin{barticle}[auto:STB|2013/06/05|13:45:01]
\bauthor{\bsnm{Decelle},~\bfnm{A.}\binits{A.}},
\bauthor{\bsnm{Krzakala},~\bfnm{F.}\binits{F.}},
\bauthor{\bsnm{Moore},~\bfnm{C.}\binits{C.}} \AND
\bauthor{\bsnm{Zdeborov{\'a}},~\bfnm{L.}\binits{L.}}
(\byear{2012}).
\btitle{Asymptotic analysis of the stochastic block model for modular networks
and its algorithmic applications}.
\bjournal{Phys. Rev. E (3)}
\bvolume{84}
\bpages{066106}.
\bptok{imsref}%
\end{barticle}
\endbibitem

\bibitem{Fortunato2010}
\begin{barticle}[mr]
\bauthor{\bsnm{Fortunato},~\bfnm{Santo}\binits{S.}}
(\byear{2010}).
\btitle{Community detection in graphs}.
\bjournal{Phys. Rep.}
\bvolume{486}
\bpages{75--174}.
\bid{doi={10.1016/j.physrep.2009.11.002}, issn={0370-1573}, mr={2580414}}
\bptok{imsref}%
\end{barticle}
\endbibitem

\bibitem{Gle75}
\begin{barticle}[mr]
\bauthor{\bsnm{Gleser},~\bfnm{Leon~Jay}\binits{L.~J.}}
(\byear{1975}).
\btitle{On the distribution of the number of successes in independent trials}.
\bjournal{Ann. Probab.}
\bvolume{3}
\bpages{182--188}.
\bid{mr={0365651}}
\bptok{imsref}%
\end{barticle}
\endbibitem

\bibitem{Handcock2007}
\begin{barticle}[mr]
\bauthor{\bsnm{Handcock},~\bfnm{Mark~S.}\binits{M.~S.}},
\bauthor{\bsnm{Raftery},~\bfnm{Adrian~E.}\binits{A.~E.}} \AND
\bauthor{\bsnm{Tantrum},~\bfnm{Jeremy~M.}\binits{J.~M.}}
(\byear{2007}).
\btitle{Model-based clustering for social networks}.
\bjournal{J. Roy. Statist. Soc. Ser. A}
\bvolume{170}
\bpages{301--354}.
\bid{doi={10.1111/j.1467-985X.2007.00471.x}, issn={0964-1998}, mr={2364300}}
\bptok{imsref}%
\end{barticle}
\endbibitem

\bibitem{Hoe56}
\begin{barticle}[mr]
\bauthor{\bsnm{Hoeffding},~\bfnm{Wassily}\binits{W.}}
(\byear{1956}).
\btitle{On the distribution of the number of successes in independent trials}.
\bjournal{Ann. Math. Statist.}
\bvolume{27}
\bpages{713--721}.
\bid{issn={0003-4851}, mr={0080391}}
\bptok{imsref}%
\end{barticle}
\endbibitem

\bibitem{Hoff2007}
\begin{bincollection}[auto:STB|2013/06/05|13:45:01]
\bauthor{\bsnm{Hoff},~\bfnm{P.~D.}\binits{P.~D.}}
(\byear{2007}).
\btitle{Modeling homophily and stochastic equivalence in symmetric relational
data}.
In \bbooktitle{Advances in Neural Information Processing Systems,
Vol. 19}.
\bpublisher{MIT Press}, \blocation{Cambridge, MA}.
\bptok{imsref}%
\end{bincollection}
\endbibitem

\bibitem{Holland83}
\begin{barticle}[mr]
\bauthor{\bsnm{Holland},~\bfnm{Paul~W.}\binits{P.~W.}},
\bauthor{\bsnm{Laskey},~\bfnm{Kathryn~Blackmond}\binits{K.~B.}} \AND
\bauthor{\bsnm{Leinhardt},~\bfnm{Samuel}\binits{S.}}
(\byear{1983}).
\btitle{Stochastic blockmodels: First steps}.
\bjournal{Social Networks}
\bvolume{5}
\bpages{109--137}.
\bid{doi={10.1016/0378-8733(83)90021-7}, issn={0378-8733}, mr={0718088}}
\bptok{imsref}%
\end{barticle}
\endbibitem

\bibitem{HolLei81}
\begin{barticle}[mr]
\bauthor{\bsnm{Holland},~\bfnm{Paul~W.}\binits{P.~W.}} \AND
\bauthor{\bsnm{Leinhardt},~\bfnm{Samuel}\binits{S.}}
(\byear{1981}).
\btitle{An exponential family of probability distributions for directed
graphs}.
\bjournal{J. Amer. Statist. Assoc.}
\bvolume{76}
\bpages{33--65}.
\bid{issn={0162-1459}, mr={0608176}}
\bptnote{check related}%
\bptok{imsref}%
\end{barticle}
\endbibitem

\bibitem{Karrer10}
\begin{barticle}[mr]
\bauthor{\bsnm{Karrer},~\bfnm{Brian}\binits{B.}} \AND
\bauthor{\bsnm{Newman},~\bfnm{M.~E.~J.}\binits{M.~E.~J.}}
(\byear{2011}).
\btitle{Stochastic blockmodels and community structure in networks}.
\bjournal{Phys. Rev. E (3)}
\bvolume{83}
\bpages{016107, 10}.
\bid{doi={10.1103/PhysRevE.83.016107}, issn={1539-3755}, mr={2788206}}
\bptok{imsref}%
\end{barticle}
\endbibitem

\bibitem{Mariadassouetal2010}
\begin{barticle}[mr]
\bauthor{\bsnm{Mariadassou},~\bfnm{Mahendra}\binits{M.}},
\bauthor{\bsnm{Robin},~\bfnm{St{\'e}phane}\binits{S.}} \AND
\bauthor{\bsnm{Vacher},~\bfnm{Corinne}\binits{C.}}
(\byear{2010}).
\btitle{Uncovering latent structure in valued graphs: A variational approach}.
\bjournal{Ann. Appl. Stat.}
\bvolume{4}
\bpages{715--742}.
\bid{doi={10.1214/10-AOAS361}, issn={1932-6157}, mr={2758646}}
\bptok{imsref}%
\end{barticle}
\endbibitem

\bibitem{Mosseletal2012}
\begin{bmisc}[auto:STB|2013/06/05|13:45:01]
\bauthor{\bsnm{Mossel},~\bfnm{E.}\binits{E.}},
\bauthor{\bsnm{Neeman},~\bfnm{J.}\binits{J.}} \AND
\bauthor{\bsnm{Sly},~\bfnm{A.}\binits{A.}}
(\byear{2012}).
\bhowpublished{Stochastic block models and reconstruction. Available at
\arxivurl{arXiv:1202.1499}}.
\bptok{imsref}%
\end{bmisc}
\endbibitem

\bibitem{Newman2004Review}
\begin{barticle}[auto:STB|2013/06/05|13:45:01]
\bauthor{\bsnm{Newman},~\bfnm{M.~E.~J.}\binits{M.~E.~J.}}
(\byear{2004}).
\btitle{Detecting community structure in networks}.
\bjournal{Eur. Phys. J. B}
\bvolume{38}
\bpages{321--330}.
\bptok{imsref}%
\end{barticle}
\endbibitem

\bibitem{Newman2006}
\begin{barticle}[mr]
\bauthor{\bsnm{Newman},~\bfnm{M.~E.~J.}\binits{M.~E.~J.}}
(\byear{2006}).
\btitle{Finding community structure in networks using the eigenvectors of
matrices}.
\bjournal{Phys. Rev. E (3)}
\bvolume{74}
\bpages{036104, 19}.
\bid{doi={10.1103/PhysRevE.74.036104}, issn={1539-3755}, mr={2282139}}
\bptok{imsref}%
\end{barticle}
\endbibitem

\bibitem{NewmanPNAS}
\begin{barticle}[auto:STB|2013/06/05|13:45:01]
\bauthor{\bsnm{Newman},~\bfnm{M.~E.~J.}\binits{M.~E.~J.}}
(\byear{2006}).
\btitle{Modularity and community structure in networks}.
\bjournal{Proc. Natl. Acad. Sci. USA}
\bvolume{103}
\bpages{8577--8582}.
\bptok{imsref}%
\end{barticle}
\endbibitem

\bibitem{Newman&Girvan2004}
\begin{barticle}[auto:STB|2013/06/05|13:45:01]
\bauthor{\bsnm{Newman},~\bfnm{M.~E.~J.}\binits{M.~E.~J.}} \AND
\bauthor{\bsnm{Girvan},~\bfnm{M.}\binits{M.}}
(\byear{2004}).
\btitle{Finding and evaluating community structure in networks}.
\bjournal{Phys. Rev. E (3)}
\bvolume{69}
\bpages{026113}.
\bptok{imsref}%
\end{barticle}
\endbibitem

\bibitem{Newman&Leicht2007}
\begin{barticle}[auto:STB|2013/06/05|13:45:01]
\bauthor{\bsnm{Newman},~\bfnm{M.~E.~J.}\binits{M.~E.~J.}} \AND
\bauthor{\bsnm{Leicht},~\bfnm{E.~A.}\binits{E.~A.}}
(\byear{2007}).
\btitle{Mixture models and exploratory analysis in networks}.
\bjournal{Proc. Natl. Acad. Sci. USA}
\bvolume{104}
\bpages{9564--9569}.
\bptok{imsref}%
\end{barticle}
\endbibitem

\bibitem{Nowicki2001}
\begin{barticle}[mr]
\bauthor{\bsnm{Nowicki},~\bfnm{Krzysztof}\binits{K.}} \AND
\bauthor{\bsnm{Snijders},~\bfnm{Tom A.~B.}\binits{T.~A.~B.}}
(\byear{2001}).
\btitle{Estimation and prediction for stochastic blockstructures}.
\bjournal{J. Amer. Statist. Assoc.}
\bvolume{96}
\bpages{1077--1087}.
\bid{doi={10.1198/016214501753208735}, issn={0162-1459}, mr={1947255}}
\bptok{imsref}%
\end{barticle}
\endbibitem

\bibitem{Perry2012}
\begin{bmisc}[auto:STB|2013/06/05|13:45:01]
\bauthor{\bsnm{Perry},~\bfnm{P.~O.}\binits{P.~O.}} \AND
\bauthor{\bsnm{Wolfe},~\bfnm{P.~J.}\binits{P.~J.}}
(\byear{2012}).
\bhowpublished{Null models for network data. Available at
\arxivurl{arXiv:1201.5871v1}}.
\bptok{imsref}%
\end{bmisc}
\endbibitem

\bibitem{Rohe2011}
\begin{barticle}[mr]
\bauthor{\bsnm{Rohe},~\bfnm{Karl}\binits{K.}},
\bauthor{\bsnm{Chatterjee},~\bfnm{Sourav}\binits{S.}} \AND
\bauthor{\bsnm{Yu},~\bfnm{Bin}\binits{B.}}
(\byear{2011}).
\btitle{Spectral clustering and the high-dimensional stochastic blockmodel}.
\bjournal{Ann. Statist.}
\bvolume{39}
\bpages{1878--1915}.
\bid{doi={10.1214/11-AOS887}, issn={0090-5364}, mr={2893856}}
\bptok{imsref}%
\end{barticle}
\endbibitem

\bibitem{Shi00}
\begin{barticle}[auto:STB|2013/06/05|13:45:01]
\bauthor{\bsnm{Shi},~\bfnm{J.}\binits{J.}} \AND
\bauthor{\bsnm{Malik},~\bfnm{J.}\binits{J.}}
(\byear{2000}).
\btitle{Normalized cuts and image segmentation}.
\bjournal{IEEE Trans. Pattern Analysis and Machine Intelligence}
\bvolume{22}
\bpages{888--905}.
\bptok{imsref}%
\end{barticle}
\endbibitem

\bibitem{Snijders&Nowicki1997}
\begin{barticle}[mr]
\bauthor{\bsnm{Snijders},~\bfnm{Tom A.~B.}\binits{T.~A.~B.}} \AND
\bauthor{\bsnm{Nowicki},~\bfnm{Krzysztof}\binits{K.}}
(\byear{1997}).
\btitle{Estimation and prediction for stochastic blockmodels for graphs with
latent block structure}.
\bjournal{J. Classification}
\bvolume{14}
\bpages{75--100}.
\bid{doi={10.1007/s003579900004}, issn={0176-4268}, mr={1449742}}
\bptok{imsref}%
\end{barticle}
\endbibitem

\bibitem{wang1987}
\begin{barticle}[mr]
\bauthor{\bsnm{Wang},~\bfnm{Yuchung~J.}\binits{Y.~J.}} \AND
\bauthor{\bsnm{Wong},~\bfnm{George~Y.}\binits{G.~Y.}}
(\byear{1987}).
\btitle{Stochastic blockmodels for directed graphs}.
\bjournal{J. Amer. Statist. Assoc.}
\bvolume{82}
\bpages{8--19}.
\bid{issn={0162-1459}, mr={0883333}}
\bptok{imsref}%
\end{barticle}
\endbibitem

\bibitem{Wu1983}
\begin{barticle}[mr]
\bauthor{\bsnm{Wu},~\bfnm{C.~F.~Jeff}\binits{C.~F.~J.}}
(\byear{1983}).
\btitle{On the convergence properties of the {EM} algorithm}.
\bjournal{Ann. Statist.}
\bvolume{11}
\bpages{95--103}.
\bid{doi={10.1214/aos/1176346060}, issn={0090-5364}, mr={0684867}}
\bptok{imsref}%
\end{barticle}
\endbibitem

\bibitem{Yao03}
\begin{bincollection}[auto:STB|2013/06/05|13:45:01]
\bauthor{\bsnm{Yao},~\bfnm{Y.~Y.}\binits{Y.~Y.}}
(\byear{2003}).
\btitle{Information-theoretic measures for knowledge discovery and data
mining}.
In \bbooktitle{Entropy Measures, Maximum Entropy Principle and Emerging
Applications}
\bpages{115--136}.
\bpublisher{Springer}, \blocation{New York}.
\bptok{imsref}%
\end{bincollection}
\endbibitem

\bibitem{Zhaoetal2012}
\begin{barticle}[mr]
\bauthor{\bsnm{Zhao},~\bfnm{Yunpeng}\binits{Y.}},
\bauthor{\bsnm{Levina},~\bfnm{Elizaveta}\binits{E.}} \AND
\bauthor{\bsnm{Zhu},~\bfnm{Ji}\binits{J.}}
(\byear{2012}).
\btitle{Consistency of community detection in networks under degree-corrected
stochastic block models}.
\bjournal{Ann. Statist.}
\bvolume{40}
\bpages{2266--2292}.
\bid{issn={0090-5364}, mr={3059083}}
\bptok{imsref}%
\end{barticle}
\endbibitem

\end{thebibliography}
\end{document}